\documentclass[trackchanges,twocolumn,tighten]{aastex62} 

\usepackage{amsmath}

\usepackage{longtable}
\usepackage{soul}
\usepackage{CJK}
\usepackage{lineno}

\newcommand{\Msun}{M_\odot}

\newcommand{\Teff}{T_\mathrm{eff}}

\newcommand{\bbud}{\alpha\;\mathrm{Ori~B}}

\definecolor{jaredpurple}{RGB}{93, 63, 211}
\definecolor{meridithgreen}{RGB}{0, 150, 0}
\definecolor{annablue}{RGB}{29, 162, 219}

\newcommand{\gtapprox}{\gtrsim}
\newcommand{\ltapprox}{\lesssim}

\newcommand{\appropto}{\mathrel{\vcenter{
		\offinterlineskip\halign{\hfil$##$\cr
	\propto\cr\noalign{\kern2pt}\sim\cr\noalign{\kern-2pt}}}}}

\newlength{\apjcolwidth}
\newlength{\figwidth}
\newlength{\doublewide}
\begin{document}
\title{Betelgeuse, Betelgeuse, Betelgeuse, Betel-buddy? Constraints on the dynamical companion to $\alpha$~Orionis from HST}

\shorttitle{Hunting for $\bbud$ with HST}
\shortauthors{Goldberg, O'Grady, Joyce et al.}


\author[0000-0003-1012-3031]{Jared A. Goldberg}
\affiliation{Center for Computational Astrophysics, Flatiron Institute, New York, NY, USA}

\author[0000-0002-7296-6547]{Anna J. G. O'Grady}
\altaffiliation{McWilliams Fellow}
\affiliation{McWilliams Center for Cosmology and Astrophysics, Department of Physics, Carnegie Mellon University, Pittsburgh, PA 15213, USA}

\author[0000-0002-8717-127X]{Meridith Joyce}
\affiliation{University of Wyoming, 1000 E University Ave, Laramie, WY USA}
\affiliation{Konkoly Observatory, HUN-REN CSFK, Konkoly-Thege Mikl\'os \'ut 15-17, H-1121, Budapest, Hungary}
\affiliation{CSFK, MTA Centre of Excellence, Konkoly-Thege Mikl\'os \'ut 15-17, H-1121, Budapest, Hungary}

\author[0000-0002-8878-3315]{Christian I. Johnson}
\affiliation{Space Telescope Science Institute
3700 San Martin Drive
Baltimore, MD 21218, USA}

\author[0000-0002-8159-1599]{L\'{a}szl\'{o} Moln\'{a}r}
\affiliation{Konkoly Observatory, HUN-REN CSFK, Konkoly-Thege Mikl\'os \'ut 15-17, H-1121, Budapest, Hungary}
\affiliation{CSFK, MTA Centre of Excellence, Konkoly-Thege Mikl\'os \'ut 15-17, H-1121, Budapest, Hungary}
\affiliation{E\"otv\"os Lor\'and University, Institute of Physics and Astronomy, P\'azm\'any P\'eter s\'et\'any 1/A, H-1117, Budapest, Hungary}

\author[0000-0002-8985-8489]{Andrea K. Dupree}
\affiliation{Center for Astrophysics $|$ Harvard \& Smithsonian 60 Garden Street Cambridge, MA 02138, USA}

\author[0000-0002-9700-0036]{Brendan O'Connor}
\altaffiliation{McWilliams Fellow}
\affiliation{McWilliams Center for Cosmology and Astrophysics, Department of Physics, Carnegie Mellon University, Pittsburgh, PA 15213, USA}

\author[0000-0001-7081-0082]{Maria R. Drout}
\affiliation{David A. Dunlap Department of Astronomy \& Astrophysics, University of Toronto, 50 St. George Street, Toronto, Ontario, M5S 3H4, Canada}

\author[0000-0002-0870-6388]{Maxwell Moe}
 \affiliation{University of Wyoming, 1000 E University Ave, Laramie, WY USA}

\author[0000-0001-5228-6598]{Katelyn Breivik}
\affiliation{McWilliams Center for Cosmology and Astrophysics, Department of Physics, Carnegie Mellon University, Pittsburgh, PA 15213, USA}

\author[0000-0002-0882-7702]{Annalisa Calamida}
 \affiliation{Space Telescope Science Institute, 3600 San Martin Drive, Baltimore, 21218 MD, USA}
 \affiliation{INAF - Osservatorio Astronomico Capodimonte, Salita Moiariello 16, 80131 Napoli, Italy}

\author[0009-0008-1688-2323]{Iman Behbehani}
\affiliation{City University of New York, Graduate Center 365 5th Ave, 10006, New York, NY, USA}

\author[0000-0002-3780-0592]{Niall J. Miller}
\affiliation{University of Wyoming, 1000 E University Ave, Laramie, WY USA}

\correspondingauthor{J.~A.~Goldberg}
\email{jgoldberg@flatironinstitute.org}

\begin{abstract}
Recently, two independent analyses have asserted that the cause of the Long Secondary Period (LSP) observed in the variability spectrum of our nearest red supergiant, Betelgeuse ($\alpha$ Ori), is an as-yet undetected, low-mass binary companion dubbed $\bbud$. In this paper, we present the results of a far-UV observational campaign using the 
STIS echelle spectrograph on the Hubble Space Telescope aimed at detecting spectral signatures of the companion. The four-quadrant tiling pattern and timing of the observations were optimized to isolate the companion, with observations taking place during a period of maximum angular and velocity separation between Betelgeuse and the putative companion. Spectral differencing between quadrants recovers no spectral features at the companion's velocity in excess of the background or Betelgeuse's chromosphere, i.e. a non-detection. 
Having determined that $\bbud$ is most likely a Young Stellar Object (YSO) thanks to constraints from a complementary X-ray campaign with the Chandra X-ray Observatory in a companion paper, comparison of our data against canonical spectra from YSOs in the ULLYSES database allows us to confidently exclude masses above $\gtrsim1.5M_\odot$ and companion continuum or line emission in excess of $\approx10^{-14}$\,erg\,s$^{-1}$\,cm$^{-2}$\,\AA$^{-1}$ in the FUV ($\approx$1200-1700\AA). Future observational campaigns aware of the LSP phase are needed to place deeper constraints on the spectroscopic nature of $\bbud$.
\end{abstract}

\keywords{Betelgeuse, Variability, Chromospheres, Binarity, UV Spectra}

%
%
\section{Introduction}
\label{sec:intro}
Alpha Orionis ($\alpha$~Ori), also known as Betelgeuse, is the 10th brightest star in the sky \citep{Hoffleit1982, Hoffleit1987} and arguably our nearest red supergiant (RSG), with modern distance estimates ranging from 168\,pc \citep{Joyce2020} to 222\,pc \citep{Harper2017} from Earth. Interest in this star has increased significantly over the past five years due in large part to a severe brightness drop 
of $V\sim1.6$\,mag in 2019 that has come to be known as ``The Great Dimming'' \citep[e.g.][]{Dupree-2020,Dharmawardena2020,Levesque2020,Montarges2021,Harper-SOFIA-2020,Harper2020,Kravchenko2021,Taniguchi2022,Cannon-2023,Jadlovsky2024,Drevon2024}. Given that Betelgeuse is an evolved massive star ($M\approx18M_\odot$, e.g. \citealt{Joyce2020}), this event prompted speculation that the star was close to undergoing a supernova (SN) explosion, but investigation suggests that Betelgeuse's behavior during the Great Dimming comes from the creation of a dust cloud and a subsequent mode transition \citep{Levesque2020,Montarges2021,Dupree2022,MacLeod2023}. These phenomena occur in the stellar and circumstellar envelope, and therefore need not be connected to any late-stage core activity that would suggest an imminent death. 
Current consensus favors an SN being several hundred thousand years away \citep[e.g.][]{Joyce2020, Neuhauser2022, MacLeod2023}, but tension in the literature nonetheless remains \citep{Saio2023}.

At the heart of the argument for or against an imminent SN is the classification of Betelgeuse's periodicities and their driving mechanisms \citep{Joyce2020, MacLeod2023}. Betelgeuse exhibits a number of strong periodic signals, the most prominent of which are periodicities at roughly 2100 days, 400 days, and 185 days. 
If the $\approx$2100-day periodicity is an \textit{intrinsic} variation corresponding to Betelgeuse's fundamental (acoustic) pressure mode (FM), this would imply a radius in excess of $1000\,R_\odot$ and place the star in its core carbon fusion stage of evolution: a short-lived phase in the star's last few hundred to few thousand years of life \citep{Saio2023}. The consensus regarding mode identification, however, is that the $\approx$400--day periodicity is the FM \citep{Kiss2006,Joyce2020, Dupree2022, Neuhauser2022, Granzer2022, MacLeod2023, MolnarResearchNote} which represents the slowest (lowest-frequency) form of intrinsic variability caused by acoustic pressure variation.

This identification, coupled with other high-fidelity constraints, requires that Betelgeuse is undergoing He fusion in the core and therefore has $\sim10^5$\,yr before its eventual SN. However, it also leaves open the question of what drives the longer (even lower-frequency), $\approx$2100--day periodicity. 
Long periodicities occurring at sub-FM frequencies are observed in about one third of cool, luminous stars with large, radially extended convective envelopes \citep[e.g.][]{Payne1954,Wood1999,Wood2004,Nicholls2009,Soszynski2007,soszynski2009-lmc,soszynski2011-smc,soszynski2013-bulge,Soszynski2021,Pawlak2021,Pawlak2023}, including Red Supergiants such as Betelgeuse \citep[e.g.][]{Kiss2006,Yang2012,Percy2014,Chatys2019}. 

These sub-FM periodicities have been given the name ``Long Secondary Periods,'' or LSPs.    
There have been many mechanisms proposed to explain LSPs in general, ranging from giant cell convection \citep[e.g.][]{Stothers1972,Stothers2010}, mode interactions \citep[e.g.][]{Saio2015}, non-radial pulsations \citep[e.g.][]{Wood2000}, and dust modulation beyond the stellar surface \citep[e.g.][]{Wood2004,Wood2009,Percy2023}. The most salient hypothesis, however, is binarity: namely, that the long secondary period is induced by an orbital companion (e.g. \citealt[][]{Wood2004,Derekas2006,Soszynski2007,Soszynski2021}). 

Recently, \citet{Goldberg2024} and \citet{MacLeod2025} independently proposed the existence of a low-mass companion in a $\approx2100$\,d orbit as the explanation for the LSP observed in Betelgeuse, which is present in both the star's lightcurve and its spectroscopic radial velocity (RV). Through a joint analysis of Betelgeuse's brightness and RV variations and the phase offset between them compared to a sample of other stars with measured LSPs and RVs, \citet{Goldberg2024} argued that the binarity hypothesis is the best explanation for Betelgeuse's LSP and RV signal, systematically ruling out all other proposed causes of the LSP in Betelgeuse. 
\citet{MacLeod2025} infer the existence of an $\bbud$ using a long baseline of RV data as well as astrometric analysis, finding a periodic signal consistent with orbital motion on the LSP timescale in both data sets. In both of these works, the LSP is attributed to a low-mass ($0.5-2.0\,M_\odot$) binary companion on a $\approx6$\,yr, $\approx9$\,AU, low-eccentricity orbit, at relatively low inclination relative to the observer. A companion at such a low mass and wide orbital separation would not directly affect Betelgeuse's evolution through any substantial mass transfer, but it may interact indirectly via tides or by accreting from Betelgeuse's stellar wind.

The idea that Betelgeuse could presently be a binary system has been explored in the literature. 
For example, from speckle-imaging measurements \citet{Karovska1986} proposed that Betelgeuse has two high-eccentricity companions, including one at a separation of 0.06" (near that of $\bbud$) but a period of 2.1 years.\footnote{This particular configuration was not confirmed in follow-up work with higher-resolution instruments \citep[e.g.][]{Wilson1992, Kervella2009,Montarges2016}, nor recovered in the lightcurve or RV signal \citep{Joyce2020,Goldberg2024,MacLeod2023,MacLeod2025}.}
Moreover, a companion orbiting Betelgeuse is consistent with expectations for stars in Betelgeuse's mass class ($>10\Msun$).
Betelgeuse evolved from a late-O/early-B, ${\approx}18 M_\odot$ main-sequence (MS) star. The binary fraction of late-O/early-B stars is 95\% \citep[e.g.][]{Sana2012,deMink2013}. Though close binaries ($a\lesssim0.5$ AU, which would have interacted by the RSG phase) show a uniform distribution of mass ratio $(q=M_2/M_1)$ with an excess at $q\approx1$, at wider separations the mass ratio companions are skewed toward small mass ratios, favoring $q<0.2$ at $\approx$10AU \citep{Rizzuto2013, Moe2017, Offner2023}. 
Other systems have been confirmed to have companions with low $q$ and intermediate separations, such as a large population of Cepheids in the $5-10\Msun$ range with binary companions on 1-20 yr periods, identified via both RV measurements \citep[e.g.][]{Evans2015} and X-ray emission and UV excess \citep{Evans2022}.
It is therefore unsurprising that such a massive star has a $q\ltapprox0.05$ companion at $8.6 \pm 0.3$\,AU.

Moving beyond dynamical arguments, observational follow-up informed by the companion's orbital configuration will further constrain its nature and impact on the star and its environment.
Importantly, \citet{Goldberg2024} show that the phase offset between the RV and light curves requires the companion to be \textit{behind} Betelgeuse during Betelgeuse's LSP brightness minima. The Great Dimming, which prompted many immediate follow-up observations, occurred at the brightness minimum of both the stellar pulsation and the LSP. However, at the time the companion was \textit{behind} the disc of Betelgeuse, so observations then 
\citep[e.g.][]{Dupree-2020,Dupree2022,Kashyap-2020,Dharmawardena2020,Levesque2020,Montarges2021,Kravchenko2021,Alexeeva2021,Matthews2022,Taniguchi2022}
would not show any emission from the companion. Epochs corresponding to other orbital configurations, on the other hand, are ripe for further exploration.

\citet{MacLeod2025} further argue that the companion could plausibly induce rotational velocities of $\approx5-15$km/s via tidal interactions, 
matching the measured 36 year period in the lightcurve and RV \citep{Joyce2020,MacLeod2025} as well as observations of a dipolar ``rotational" velocity field in spatially-resolved observations \citep[redshifted in the SE and blueshifted in the NW, see, e.g.][]{Dupree1987,Gilliland1996,Uitenbroek1998,Lobel-2001,Kervella2009,Kervella2018}. Fast rotation in Betelgeuse has previously been argued to indicate a recent merger, as rapid rotation of a RSG at the km/s level requires an external source of angular momentum input into the star's outer envelope \citep[e.g.][]{Wheeler2017,Wheeler2023,Chatzopoulos2020,Nance2018,Shiber2024}.\footnote{See, however, discussion by \citet{Ma2024} on under-resolved large-scale convection as a plausible origin for the \citet{Kervella2018} velocity field, a counter-hypothesis which predicts a dipole orientation which changes stochastically and decoheres on a $\lesssim$ year-long convective overturn timescale.} 
If the companion is instead responsible, then when combined with an orbital phase inferred from Betelgeuse's RV, the direction of rotation (PA of 48$\pm3.5^\circ$ W. of S.; \citealt{Kervella2018}) would constrain the companion's projected orbital plane and thus on-sky location relative to Betelgeuse.

We obtained Director's Discretionary Time (DDT) during Hubble Space Telescope (HST) Cycle 32 to observe Betelgeuse and its putative companion with the Space Telescope Imaging Spectrograph (STIS) instrument in the Far-UV (FUV) in a configuration never before observed. Though \citet{MacLeod2025} argue that the chromospheric emission from Betelgeuse would prevent detection of a $\approx0.5\,M_{\odot}$ companion, the mass range---which extends potentially up to $2\,M_{\odot}$ permitted by the high-quality STELLA RVs \citep{Goldberg2024} and consistent with the astrometric analysis of \citet{MacLeod2025}---includes a number of scenarios that could be detectable in both the far-UV and X-ray bands (see our companion publication in \citealt{BetelChandra2025}). 

This work is organized as follows: In Section \ref{sec:expectations}, we discuss motivations for our choices to observe the Betelgeuse system in the FUV and at quadrature. Section~\ref{sec:observations} then describes our HST configuration, tiling pattern, and observational timing. Section~\ref{sec:analysis} presents our spectroscopic observations, and discusses the constraints we can place on the FUV emission from $\bbud$. We summarize and conclude in Section~\ref{sec:conclusions}.

\section{Expectations for UV emission}
\label{sec:expectations}
Our observing strategy is informed by the various mass estimates and constraints for $\bbud$, and the most likely evolutionary possibilities for its companion.  In this section we discuss the range of expectations for the companion's spectral properties, and emphasize the importance of observing at $\bbud$'s peak expected offset in redshift from Betelgeuse. 
\begin{figure}
\centering
\includegraphics[width=1.05\columnwidth]{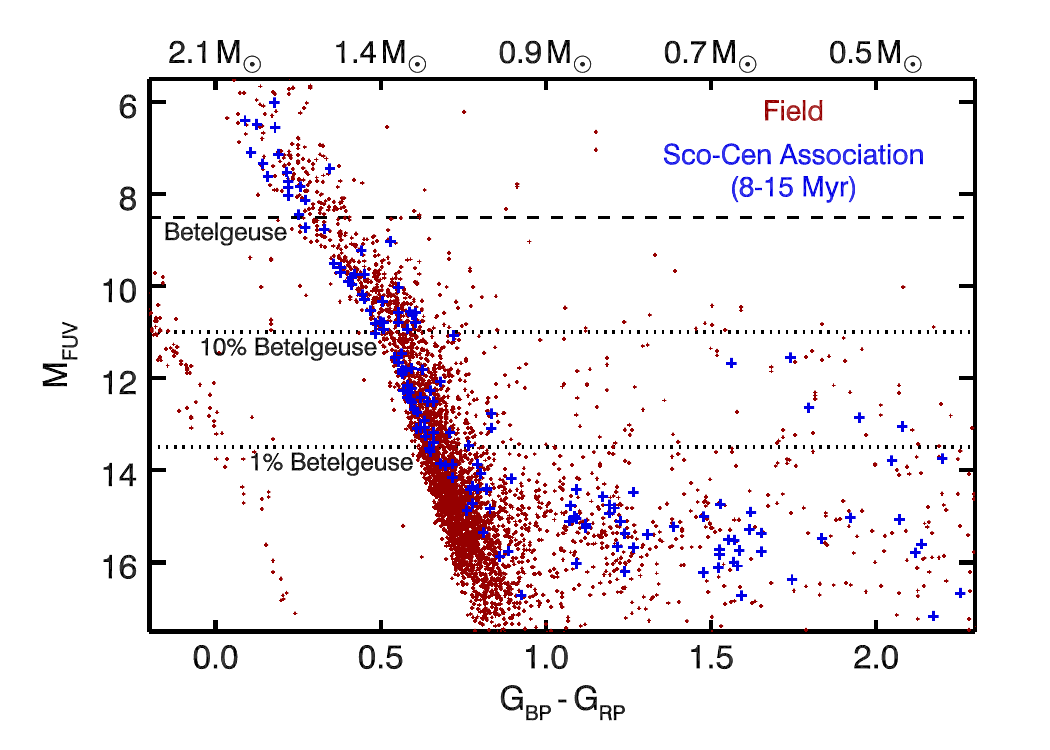}
\caption{Absolute GALEX FUV magnitudes versus Gaia color/approximate stellar mass for field stars (red) and YSOs in the Sco-Cen Association from \citet{Luhman2022} (blue). Horizontal lines indicate the FUV brightness of Betelgeuse in the GALEX band estimated from \citet{Carpenter.2018.ASTRALBetel} (dashed), well as $10\%$ and $1\%$ of Betelgeuse's brightness (dotted). Note the diversity of the YSO population, with factor-of-10 spread in the brightness for a given optical color.}
\label{fig:max}
\end{figure}

\subsection{Possible Companion Scenarios}
A mass range of $\sim$\,0.5--2 M$_{\odot}$ enables a range of possible identities for the companion. One astrophysical object satisfying the mass condition is a low-mass, ``standard'' star. 
Under co-natal and coeval assumptions for the system \citep{Dolan.M.2016.BetelEvo, Joyce2020,Neuhauser2022}, 
such a star must in fact be a young stellar object, or YSO, on the basis of Betelgeuse's age of roughly 10 Myr \citep{Tognelli2011yso,Baraffe2015yso,Haemmerle2019yso,Amard2019yso}. 
It is possible that the companion is not co-natal, but rather has been captured due to dynamical interactions, with the system then ejected from its birth cluster. However, Betelgeuse's proper motion has not yet been traced back to a specific cluster \citep[though see some discussion in][]{Harper2008,Bouy2014,Bouy2015}. Moreover, ejections generally happen within 1 Myr of cluster formation \citep{Oh2016}, but the age of Betelgeuse and all nearby clusters and OB associations are much older than 1 Myr \citep[e.g.][]{Blaauw1964,Briceno2007,Briceno2019}, so even a dynamical ejection scenario would be close to coeval.

Other possibilities include a more exotic, compact object, such as a neutron star (NS) or white dwarf (WD).
In the case of a WD, it is not possible to construct a non-interacting, two-body, coeval evolutionary scenario that combines a 10-Myr-old red supergiant with even the highest-mass WD, which has a theoretical lower age limit of 40 Myr  and a practical (observed) lower age limit of 70 Myr. 
While more complex three-body (or more) scenarios involving the deposition of a WD into Betelgeuse's orbit from some other system are theoretically possible, the kinematic signatures of such an event are unclear.
Evolutionary timescales therefore preclude the most likely versions of a WD scenario. 

Additionally, in the case of a compact object, we would expect to see a bright X-ray signature of accretion: given the orbital separation, the companion is embedded within the dusty shell surrounding Betelgeuse \citep[e.g.,][]{Haubois2019,Haubois2023}, and would interact with Betelgeuse's wind \citep{Aizu1973,Livio1986,Blodin1990,Mukai2017,Maldonado2025}. Typical X-ray luminosities for accreting WDs are $10^{30-34}$ erg s$^{-1}$ \citep{Luna2013}, and for accreting NSs are $10^{32-36}$ erg s$^{-1}$ \citep{Yungelson2019,De2022a,De2022b}.
Complementary observations with \textit{Chandra} were performed within a few weeks of the HST observations and revealed no such signature, as discussed in detail in \citet{BetelChandra2025}.

An accreting compact object is therefore ruled out on the basis of a non-detection with \textit{Chandra} over 40ks of observations: this integration time implies an upper limit of $L_\mathrm{X}\ltapprox10^{30}$erg~s$^{-1}$ for the X-ray luminosity, orders of magnitude lower than the expected luminosity of $L_\mathrm{X}\gtapprox10^{35}$erg~s$^{-1}$ for an accreting compact object (up to assumptions regarding the circumstellar hydrogen column depth). 
While a YSO would typically have an accretion disk as well, and therefore produce an X-ray signature, the wide diversity of X-ray luminosities from YSOs makes it impossible to place constraints on the stellar companion on the basis of the \textit{Chandra} observations alone. It is also likely that, given the relatively small separation between Betelgeuse and $\bbud$ and the fact that Betelgeuse was an OB star during its main sequence, any accretion disk on a YSO companion would have been irradiated away (see \ref{sec:fluxexpectations}).

Therefore, if (even nearly) co-natal and coeval with Betelgeuse, $\bbud$ within the mass parameters of either study is most likely a disk-free YSO, or pre-Main sequence (MS) star, of slightly later spectral class than its subsequent class at ZAMS.

\subsection{Motivation to Observe at Quadrature}
The luminosity ratio between Betelgeuse and $\bbud$ in the YSO scenario is expected to be of order $10^4$\nobreakdash--$10^6$, making $\bbud$ undetectable in integrated light \citep{MacLeod2025}. 
However, in the far-UV spectrum of both YSOs and RSGs, emission from the chromosphere dominates the photosphere continuum~\citep{Linsky.J.2017.ChromosphereReview}. 
HST has resolved many chromospheric lines in the spectra of YSOs \citep{Yang.H.2012.LowResTTauriUV,RomanDuval.J.2022.ULLYSES,ULLYSES2025}, and has observed the chromosphere of Betelgeuse \citep{Carpenter.2018.ASTRALBetel}. 
This leads to three possible means of detection for $\bbud$: (1) the detection of a slight UV excess relative to observations taken where the companion is not present,
(2) strong emission lines either shifted relative to, or not expected from, the chromosphere of Betelgeuse, and (3) RV variations in those UV lines over the timescale of an LSP period. Option (3) would require continuous observation over the 5.5-year duration of an LSP, which is impractical. As \citet{MacLeod2025} point out, option (1) is tenuous given intrinsic variations in Betelgeuese's bright chromosphere. 
However, if observed at a proper orbital configuration where $\bbud$'s emission is redshifted or blueshifted relative to Betelgeuse, option (2) could be promising (whether or not the companion is coeval with Betelgeuse, and whether or not it is indeed a YSO). 

We therefore target our observations such that $\bbud$'s emission is expected to be significantly redshifted relative to Betelgeuse's,  
near receding quadrature on 6 December 2024 (\citealt{Goldberg2024}, within $\approx1\%$ of the \citet{MacLeod2025} ephemeris which places quadrature slightly earlier in mid-to-late November 2024). 
We also note that the companion's next (approaching) quadrature is years down the line (late November 2027) after passing behind the disc of Betelgeuse. 

\subsection{Far UV Expectations for a YSO Companion \label{sec:fluxexpectations}}

\begin{figure*}
\centering
\includegraphics[width=0.8\textwidth]{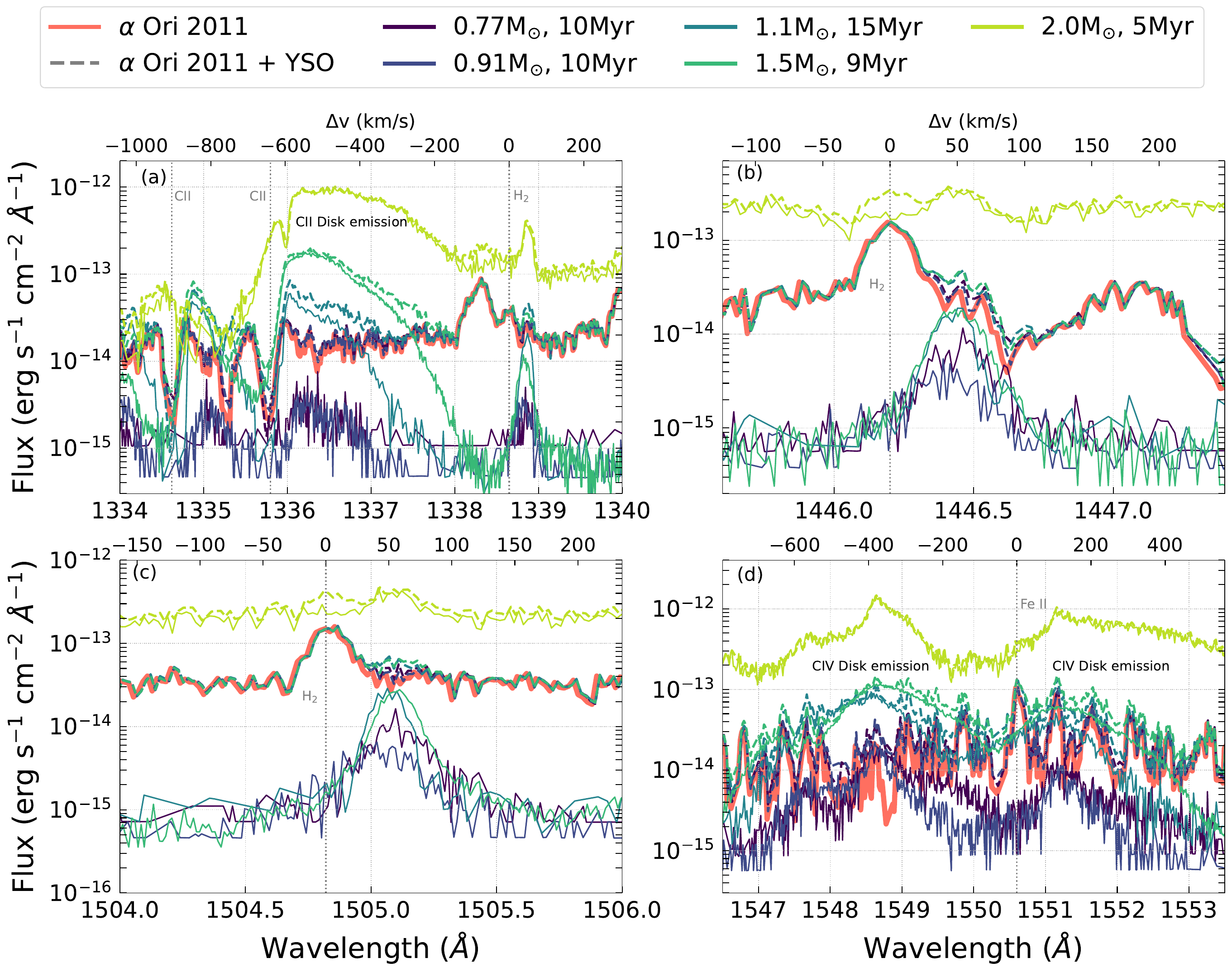}
  \caption{Examples of possibly-detectable features of $\bbud$ in the Far UV, where chromospheric emission from $\bbud$ may be stronger than, equal to, or within an order of magnitude of the emission flux from Betelgeuse for at least one example YSO. Betelgeuse data taken in 2011 as part of the ASTRAL project \citet{Carpenter.2018.ASTRALBetel} is shown as a thick orange solid line, and has been shifted to the RV of our 2024 observations. 
  YSO data from ULLYSES \citep{RomanDuval.J.2022.ULLYSES,ULLYSES2025} are shown as solid purple-blue-green lines, and have been shifted to the predicted RV of the companion's predicted 2024 receding quadrature. 
  Features of interest are indicated with vertical dotted grey lines and labeled. Wide emission features stemming from re-radiated disk emission are labeled `Disk emission' --- we do not expect this feature to be present in the spectrum of the companion. We show combined spectra (Betelgeuse + YSO) as dashed lines matching the color of the ULLYSES reference spectrum. 
  Generally, a $\gtrsim2$M$_{\odot}$ YSO should be plainly detectable, while a $\lesssim0.77$M$_{\odot}$ YSO would be undetectable. Masses between this range show varying levels of detectability depending on the precise chromospheric feature.}
     \label{fig:expectations}
\end{figure*}

In the far ultra-violet (FUV), the overall flux of Betelgeuse is expected to dominate over that of a YSO companion.
Figure \ref{fig:max} shows the Color-Magnitude Diagram (CMD) for all Gaia stars with GALEX FUV photometry ($\lambda = 1,350$--1,750\,\AA) across galactic latitudes $-15^{\circ}$--$40^{\circ}$ and longitudes $280^{\circ}$--$10^{\circ}$ within $d < 200$\,pc, which encompasses the 8--15\,Myr old Scorpius-Centaurus (Sco-Cen) OB association. The official Sco-Cen members from \citet{Luhman2022} are shown in blue. For both main sequence (MS) and $\approx$10\,Myr pre-MS stars (excluding the white dwarf sequence to the bottom left), the Gaia (BP--RP) color roughly corresponds to stellar mass, indicated on the top x-axis. 
For $M\gtrsim1.1\,M_{\odot}$ (A/F stars), the field and Sco-Cen stars follow the same MS relation. G-type ($1.0\,M_{\odot}$) and K-type ($0.8\,M_{\odot}$) stars exhibit FUV excess compared to their field MS counterparts. Early M-stars ($0.6\,M_{\odot}$) exhibit a wide range of FUV fluxes and chromospheric activity, some of them even brighter than the GK stars. This factor-of-ten spread in FUV brightness is expected, as emission lines and FUV flux depend on a star's chromospheric activity, which varies with rotation and time and environmental properties in addition to mass and age.

The weighted average FUV flux of Betelgeuse ($\approx$135--175\,nm) is $7\times 10^{-14}$\,erg\,s$^{-1}$\,cm$^{-2}$\,\AA$^{-1}$ \citep{Carpenter.2018.ASTRALBetel}, which corresponds to an apparent brightness of $m_\mathrm{FUV} = 14.65$\,mag and absolute brightness of $M_{\rm FUV} = 8.5$\,mag in the GALEX passband given the 168\,pc distance to Betelgeuse \citep{Joyce2020}. 
Horizontal lines indicating 100\%, 10\%, and 1\% of Betelgeuse's FUV brightness are shown on the CMD (Figure~\ref{fig:max}). Based on the available E140M spectrum of Betelgeuse from \citet{Carpenter.2018.ASTRALBetel}, a companion with ${\gtrsim}$30\% the FUV flux of Betelgeuse should have already been detected. 

For possible masses of $\bbud$ in the $\approx0.5-1.5M_\odot$ range, the companion's luminosity in the FUV is expected to be $>1-5\%$ of that of Betelgeuse, making it difficult to detect a shift in the overall flux level. However, the chromospheres of stars and YSOs in the Far UV are dominated by emission lines (see \citealt{Linsky.J.2017.ChromosphereReview} for a review). 
As discussed above, targeting our observations near the companion's receding quadrature corresponds to Betelgeuse's minimal RV (maximum Betelgeuse blueshift), but the companion's maximal RV (maximum $\bbud$ redshift). 
This orbital configuration entails that the emission features of $\bbud$ should be redshifted with respect to the same features of Betelgeuse, by a speed of $\approx$45.5 km/s. This is well above the detectable limit for HST-STIS with a good resolution grating ($\approx$10 km/s). Therefore, for emission lines that are sufficiently strong in both the chromospheres of Betelgeuse and YSOs, a second, redshifted emission line should be visible. 

For want of first-principles chromospheric modeling, reasonable proxies for the possible FUV spectra of $\bbud$ are provided by the ULLYSES project, which has compiled, in a publicly available database, the HST UV spectroscopy of many pre-MS/T Tauri stars \citep{RomanDuval.J.2022.ULLYSES,ULLYSES2025}. 
To illustrate this, we choose five YSOs from the ULLYSES dataset\footnote{\url{https://ullyses.stsci.edu/ullyses-targets-ttauri.html}} across the range of relevant masses and ages -- CVSO 104 (0.77M$_{\odot}$, 11 Myr), CVSO 165 (0.91M$_{\odot}$, 11 Myr), MP Mus (1.1M$_{\odot}$, 15 Myr), HD1353344B (1.5M$_{\odot}$, 9 Myr), and HD104237 (2.0M$_{\odot}$, 5 Myr). Later in this paper, we also use two other YSOs  for comparative purposes -- PDS 70 (0.76M$_{\odot}$, 11 Myr), and AK Sco (1.52M$_{\odot}$, 11Myr). The masses, distances, and $A_V$ values are obtained from the ULLYSES website and the ages from estimates of the age of the cluster containing the YSO \citep{Blaauw1964,deGeus1989age,Mamajek2002age,Muller2011age,Bossini2019age,Esposito2020age}. These objects additionally represent a wide range of $\Teff$. 

These data, along with the HST/STIS E140M spectrum of Betelgeuse from \citet{Carpenter.2018.ASTRALBetel}, are shown in Figure \ref{fig:expectations}.
At the time of the \citet{Carpenter.2018.ASTRALBetel} observations (conducted in 2011), $\bbud$ was transiting Betelgeuse, nearly at the middle of transit (see also \S~\ref{sec:timing}). 
Thus, while the overall flux of each emission line would contain contributions from both Betelgeuse and $\bbud$, their relative radial velocities should be close to zero --- the only RV shift present at that time should be the overall peculiar motion of the system ($\approx$22 km/s) as well as any RV changes induced by Betelgeuse's pulsational or convective activity. In Figure \ref{fig:expectations}, these data are shown as a thick solid orange line (labeled $\alpha$~Ori 2011), and have been blueshifted by $-2.5$ km/s to match the expected total RV of Betelgeuse during \emph{our} 2024 observations. 

We likewise shift the ULLYSES YSO spectra to be at the expected radial velocity of $\bbud$ during the 2024 observations, $\approx$45.5 km/s redshifted. We also correct for the distance of Betelgeuse (168 pc, \citealt{Joyce2020}). To account for extinction, we de-extinct the YSO spectra given their $A_V$ values from the ULLYSES project, and re-extinct them with full extinction curves given the Betelgeuse's measured extinction. 
Following \citet{Montarges2021} we adopt values of $R_V=4.2$, appropriate for supergiants \citep{Massey2005}, and $A_V=0.65$\footnote{The $A_V$ of Betelgeuse has been largely consistent across observational works. Early work by \citet{Carpenter1994} fit the GHRS-FUV spectrum with $A_V$=0.75 and $R_V$=3. \citet{Levesque2020} find $A_V$=0.62 in their 2004-2020 observations. }.  This is likely a conservative estimate, as any denser gas and dust interior to the orbit of $\bbud$, such as the inner shell at $\approx1.5R_*$ inferred from ALMA observations \citep[e.g.][]{Perrin2007, Haubois2019,Haubois2023}, would contribute to the extinction of Betelgeuse but not $\bbud$. Extinction parameters are applied to the YSO spectra for the full Milky Way extinction curve \citep{Gordon2023} using the \texttt{dust\_extinction}\footnote{\url{https://dust-extinction.readthedocs.io/en/stable/}}
package \citep{Gordon2024}. 

While the overall flux in the UV of these YSOs is only $\sim$1--5\% of that of Betelgeuse, several features could be visible in a STIS exposure containing $\bbud$ (as opposed to exposures containing only the chromosphere of Betelgeuse). There are $\approx$8 strong emission lines in the ULLYSES YSO spectra that are close to or as strong as the same line from Betelgeuse, and 4 that \emph{do not} appear as emission lines in the spectrum of Betelgeuse. 
The four panels of Figure~\ref{fig:expectations} highlight four wavelength ranges with promising features that may be visible in our observations of the system.\footnote{See Figure~7 and Table~2 of \citet{Linsky.J.2017.ChromosphereReview} for the full suite of lines we examined here and in the observing proposal.}

We also note that the ULLYSES data are all of isolated YSO systems. YSOs are often surrounded by, and accrete from, circumstellar disks, leading to the emission ``wings" seen in some of the ULLYSES spectra. These are protoplanetary disks at early ($\lesssim$10\,Myr) times, and gas-poor processed disks at later ($\gtrsim$10\,Myr) times \citep{Ribas2015}. These disks eventually photo-evaporate from high-energy radiation from the central star \citep{Gorti2009}. However, these circumstellar disks can also be photoevaporated by UV radiation from nearby OB stars \citep{Adams2004,Andrews2020}, which has been directly observed for YSOs in young OB associations \citep{Johnstone1998}. For binary YSOs, systems with circumprimary but no circumsecondary disks are also much more common than the inverse \citep{White2001}, and \citet{Kraus2012} finds that 2/3rds of YSO systems with close separations (r$\lesssim$40\,AU, compared to the $\approx$9AU separation for $\bbud$) disperse their disks with 1\,Myr of formation. Given that the $\alpha$~Ori--$\bbud$ system has a small separation relative to 40AU, and given that Betelguse was an extremely hot O star during its main-sequence lifetime, there is a high likelihood that a circumstellar disk around the companion has been dispersed by present day. Thus, the wide emission features stemming from strong winds or accretion (red/blue-shifted relative to the companion) that are common in YSO spectra (seen in the ULLYSES spectra in panels (a) and (d) of Figure~\ref{fig:expectations}), as well as wings around Lyman-$\alpha$, would almost certainly \emph{not} be present in the $\alpha$~Ori system.

We therefore focus our efforts on narrow lines without those wings, such as fluorescent H$_2$ lines (seen in all four panels of Figure~\ref{fig:expectations}). 
Of the spectra shown here, the 2\,M$_\odot$ YSO would be plainly visible in these lines, while the 0.77\,M$_\odot$ YSO would never be visible even with a large RV offset between Betelgeuse and $\bbud$. The 0.91\,M$_\odot$ YSO also appears generally too faint to detect. The 1.1 and 1.5\,M$_\odot$ YSOs are more promising to identify, though the extra flux added to the system is small.
We also note the important caveat that these are individual, empirical YSO observations, not theoretical models of YSO FUV chromospheric features across the full range of possible companion masses, for which first-principles calculations remain elusive.

\section{Observations with HST}
\label{sec:observations}

Below, we summarize the observational history of Betelgeuse to place our campaign in context. We then describe the parameters of our own observational campaign that took place in November 2024. 
We emphasize the importance of the timing of our observations relative to the companion's orbital phase (receding quadrature). We also describe our unique aperture tiling pattern, which allows for co-incident spectra which both do and do not contain emission from the companion.

\begin{figure}
\centering
\includegraphics[width=\columnwidth]{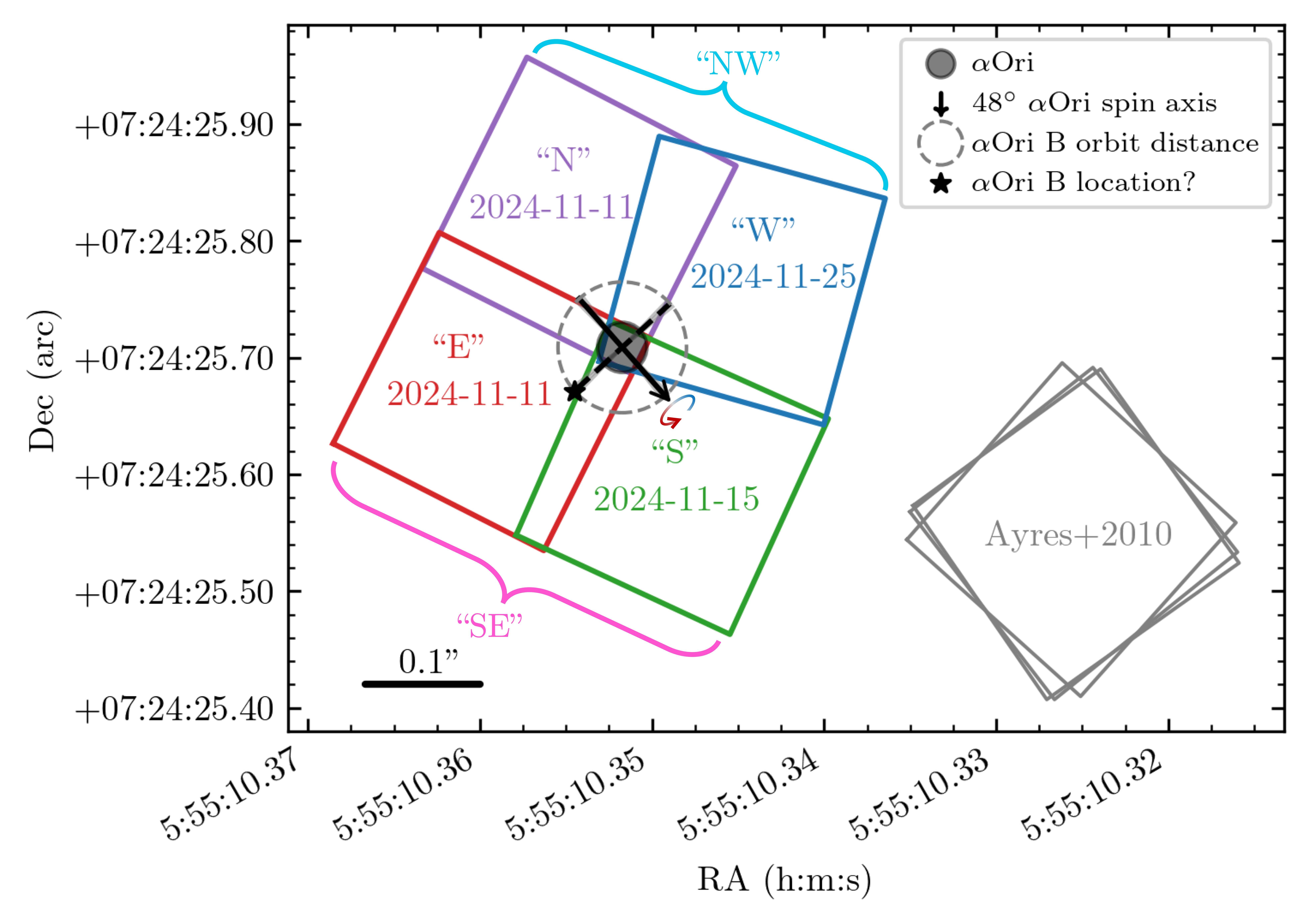}
  \caption{HST STIS E140M 0.2" aperture configuration for our observations. Our observations were taken in 4 quadrants relative to the stellar surface, with colors and labels corresponding to those used in Figs.~\ref{fig:lines_other}~and~\ref{fig:lines_lyman} and throughout the text. The ``W" quadrant (blue) is slightly tilted relative to the other quadrants due to HST's gyroscopic constraints and the slightly later observation date.
  Cyan and magenta brackets indicate our stacked quadrant groupings, labeled ``NW" and ``SE'". Betelgeuse's disc is represented by a darkgrey circle, and the downward black arrow and curved red-blue arrow indicate the observed spin axis and direction from \citet{Kervella2018} with redshifted emission to the South-East and blueshifted emission to the North-West. The orbital separation of $\bbud$ is given by the grey dashed circle, with a star denoting the approximate expected location of $\bbud$ at the time of our observations if aligned with Betelgeuse's reported plane of rotation. 
  We indicate the $0.1^"$ diameter PSF of STIS E140M with a black horizontal bar in the lower left corner.
  The patch covered by the 2011 observations (proposal: \citealt{Ayres2010}, analysis: \citealt{Carpenter.2018.ASTRALBetel}) is shown in grey; the offset is due to Betelgeuse's proper motion.}
     \label{fig:tiling}
\end{figure}

\subsection{Past UV Observations}
Long before the launch of HST, early measures of the ultraviolet spectrum of Betelgeuse (spectral type M1-M2) were obtained from a balloon-borne telescope-spectrometer, the OAO--2 spacecraft (Wisconsin photo\-meter-spectrometer), and OAO--3 (Princeton experiment on Copernicus). All detected the Mg II doublet near 2800\,\AA, a strong emission doublet in the near-ultraviolet \citep{Kondo1972,Doherty1972,Bernat-Lambert1976}. 
Subsequently, the International Ultraviolet Explorer (IUE) extended spectroscopy further into the ultraviolet to the Lyman-$\alpha$ line near 1200\,\AA{}. These low resolution spectra show no evidence of any species of ionization beyond Si II and Fe II, suggesting electron temperatures less than about 10,000\,K \citep{Basri1981}. 
The Hopkins Ultraviolet
Telescope on the Space Shuttle offered the possibility of spectroscopy from 800\,\AA\ to 1800\,\AA\ \citep{Davidsen1993}; however, the few spectra of Betelgeuse were too low signal-to-noise to be conclusive.

A substantial advance in the study of the near ultraviolet spectrum came with the Goddard High Resolution Spectrograph (GHRS) on HST, acquiring a detailed spectrum from 1980--3000\,\AA\ that enabled the identification of many features \citep{Brandt1995} from single and doubly ionized species, among them Fe II, C II, Ni II, Ti II  in addition to molecular emission of OH. The GHRS instrument also accessed the far ultraviolet spectrum, from 1200\,\AA\ to 1930\,\AA, a region that is dominated by CO bandhead absorption in addition to the singly and doubly ionized species, C I, S~I, O~I,  Cr~II, and Si II \citep{Carpenter1994}. Those authors also claim the presence of a true continuum, with a color temperature of 3500--4800\,K which they ascribe to the chromosphere, and suggest an origin in hot  chromospheric regions. A large hot region was later observed on the extended ultraviolet disk in the $\sim$2500\,\AA\ continuum, with temperatures in  excess of 200\,K from the disk value \citep{Gilliland1996}, thus supporting  origin in a hot region.  
\citet{Carpenter1994} in fact consider but dismiss the idea that a companion creates the continuum, in part because the observed flux distribution of warmer stars does not match the continuum shape.

The spectral range  below 1200\,\AA, became generally accessible with the launch of the Far Ultraviolet Spectroscopic Explorer (FUSE), which enabled access from 905--1179\,\AA\ where highly ionized species are expected. However, a spectrum of Betelgeuse revealed no sign of C III or O VI which are found in giants and supergiants of spectral type F, G, and K \citep{Dupree2005}.

More recently, observations of Betelgeuse (PI T. \citealt{Ayres2010}) were performed with STIS E140M using a square $0.2\times 0.2$\,arcsec aperture for 4.5 hours \citep{Carpenter.2018.ASTRALBetel}.The dates of these observations were February 7, 8, 24,~26, and April 1st 2011, at which point $\bbud$ was transiting (passing in front of) Betelgeuse. 
While our campaign used the same instrument, grating, and aperture as the 2011 campaign, ours differs critically in timing, tiling, and motivation. The previous \citet{Ayres2010,Carpenter.2018.ASTRALBetel} campaign was only interested in the overall UV spectrum of the chromosphere, and they did not vary their pointing. Therefore, that campaign did not obtain any spatial information and did not collect any ``clean'' spectra without $\bbud$. 

\subsection{Our FUV Observations and configuration-specific expectations}

We obtained STIS E140M spectra at four different pointings with the 0.2"$\times$0.2"
aperture, which we describe below in Section~\ref{sec:ourtiling}. 
Our campaign tiled the area including and immediately surrounding Betelgeuse into quadrants, as shown in Figure \ref{fig:tiling} with the expectation that $\bbud$ (a point source, $d\,{<}0.1$\,mas) would be in one or at most two quadrants. 
A detection would therefore localize $\alpha$ Ori B within a quadrant or hemisphere, thereby providing a constraint on the geometry of the orbit independent of inferences from Betelgeuse's apparent rotation axis.
In addition to the location of Betelgeuse and expected location of the companion if it is aligned with the \citet{Kervella2018} rotation axis, we indicate the $0.1^"$ diameter Point Spread Function (PSF) of STIS E140M as a horizontal bar. If the companion is present within the aperture of a given observation, then its observed image will be blurred over this range, with the majority of the flux remaining centrally concentrated. Flux from Betelgeuse itself, present in all 4 quadrants, will also be smeared. This provides additional motivation for quadrant-to-quadrant and hemisphere-to-hemisphere comparisons, rather than a finer-resolution search.

Calculations performed in the observing proposal demonstrated that, by subtracting two HST STIS E140M spectra, one with and one without $\bbud$, it would be possible to spot differences at the 2\% level given our proposed exposure times and anticipated S/N.
Likewise, the expected $\approx45$km/s RV shift given the system's orbital configuration is within the detectable limit for our chosen grating ($\approx$10\,km/s for E140M); therefore, for strong emission lines from both stars, we also anticipated an apparent doubling of lines with a separation of $\approx$0.16\AA\ in the exposure that contains $\bbud$ (see above discussion in \S\ref{sec:fluxexpectations}). It is also worth noting that Betelgeuse's chromospheric lines are substantially broadened by turbulence ($v_\mathrm{turb}\approx31-35$km/s, \citealt{Carpenter1997}), which is expected to exceed the broadening of the low-mass companion, but would not exceed the companion's orbital velocity. 
We predicted that our campaign would be sensitive to companions ${>}1.1M_\odot$, similar to the expected mass of $\bbud$\ recovered from the high-fidelity STELLA RV \citep{Goldberg2024} and astrometric analysis \citep{MacLeod2025}. 
We also predicted sensitivity to young, chromospherically active 0.6$M_\odot$ companions, consistent with the mass suggested by the detailed long-baseline RV analysis \citep{MacLeod2025}. 
Moreover, $\bbud$ may be accreting at a higher rate due to focused winds from Betelgeuse and thus may be even brighter in the FUV than its isolated, 10\,Myr-old counterparts in Sco-Cen. 
We therefore anticipated a reasonably good chance of detecting $\bbud$, and if not, placing constraints on remaining mass ranges allowed by a null detection. 

\subsection{HST configuration}\label{sec:ourtiling}

Observations for this project were obtained using the STIS instrument on HST via a Cycle 32 Director's Discretionary award of 4 orbits (GO/DD-17881\footnote{\href{{https://www.stsci.edu/hst-program-info/program/?program=17881}}{stsci.edu/hst-program-info/program/?program=17881}}; PI: M. Joyce).  The observations were separated into 4, 1-orbit visits in order to ease the scheduling burden introduced by the reduced gyro mode, and also to limit the number of repeat visits required should a failure occur.  The first two visits executed on 2024 November 11 with the third and fourth visits occurring on 2024 November 15 and 25, respectively.

All four visits followed the same observing plan, which utilized a 1 second imaging target acquisition with the F25ND5 attenuating filter followed by 2$\times$1000 second spectroscopic exposures using using the broad-band E140M-1425 grating setting (1150-1700 \AA) and 0.2"$\times$0.2" aperture. This choice ensured that key emission features from $\bbud$, such as Ly-$\alpha$ (1215.67 \AA), C IV (1548.2 \AA), and fluorescent H$_2$ lines (1446 \AA, 1505 \AA) would be captured.

Since the low-mass companion $\bbud$ is expected to be approximately 52\,mas from the center of Betelgeuse, we adopted the tiling pattern shown in Figure~\ref{fig:tiling}, which was constructed using combinations of Position Target (POS TARG) offset values of $\pm$0.084" 
in the X and Y coordinates. The four quadrants are labeled ``N", ``E", ``S", and ``W", loosely corresponding to their ordinal position on the sky. However, since no ORIENT constraints were applied, the roll angle of Visit 04 (the ``W" quadrant in Figure~\ref{fig:tiling}) was slightly different than the previous three visits. While the exact quadrant in which $\bbud$ is expected to appear is uncertain, the tiling strategy employed here incorporates sufficient overlap to guarantee its inclusion in at least one visit, if it is present. 

If the orbital plane of the companion is indeed perpendicular to the $48^\circ\pm3.5^\circ$ PA rotation axis \citep{Kervella2018} consistent with the rotation axis found by \citet{Uitenbroek1998,Lobel-2001}, the companion would appear in quadrant E just outside the overlap region with quadrant S, and in the SE stack. 

\subsection{Timing \label{sec:timing}}

\begin{figure}
\centering
\includegraphics[width=\columnwidth]{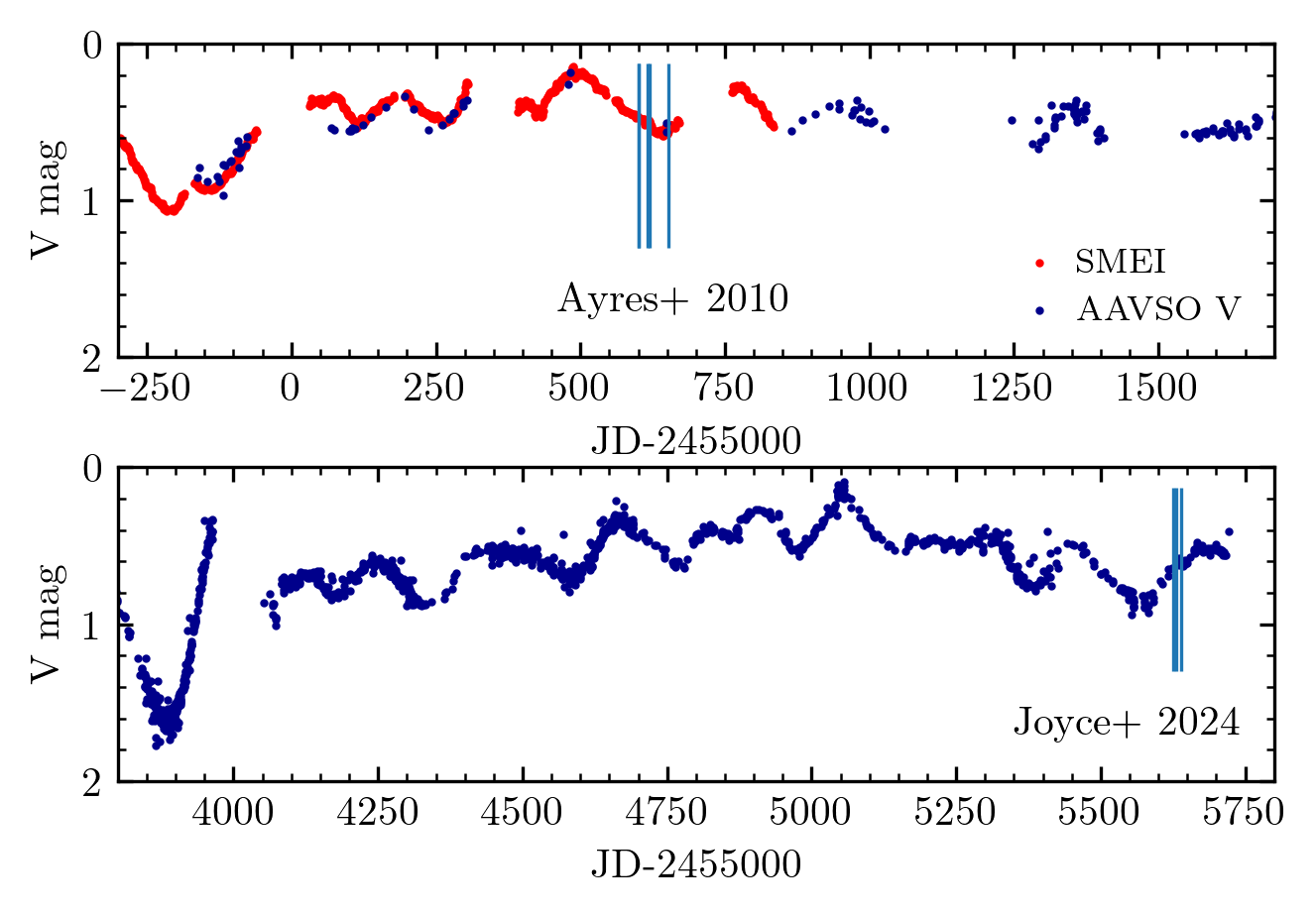}
  \caption{HST observation times relative to the photometric variations of Betelgeuse. Each panel spans just under one LSP cycle, with a cycle omitted between the two panels. 
  Red and blue points show SMEI and AAVSO \textit{V}-band photometry, as published by \citet{Joyce2020}, and extended with more recent AAVSO observations. Blue vertical lines indicate timing of the STIS E140M observations from 2011 (\citealt{Ayres2010} HST proposal, upper panel) and our 2024 observations (\citealt{Joyce2024} HST proposal, lower panel).}
     \label{fig:hst_times}
\end{figure}

\begin{figure}
\centering
\includegraphics[width=\columnwidth]{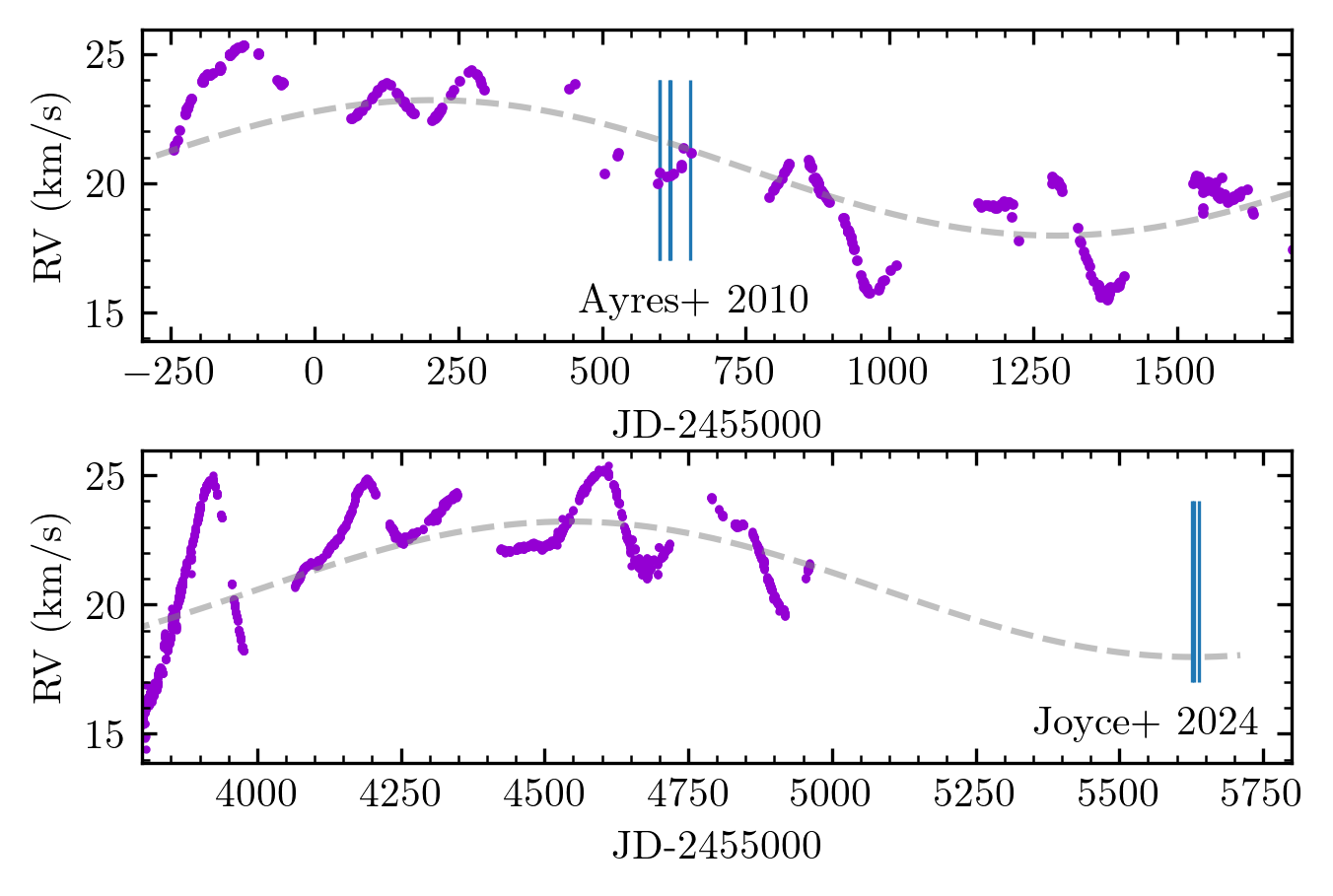}
  \caption{Same as in Figure~\ref{fig:hst_times}, but for the STELLA radial velocity measurements \citep{Granzer2022}. Since the published data ends before our new measurements, we extrapolated the LSP variation based on the sine fit shown in Figure~2 of \citet{Goldberg2024} (grey dashed line here). }
     \label{fig:hst_times_rv}
\end{figure}

\begin{figure*}
\centering
\includegraphics[width=0.75\textwidth]{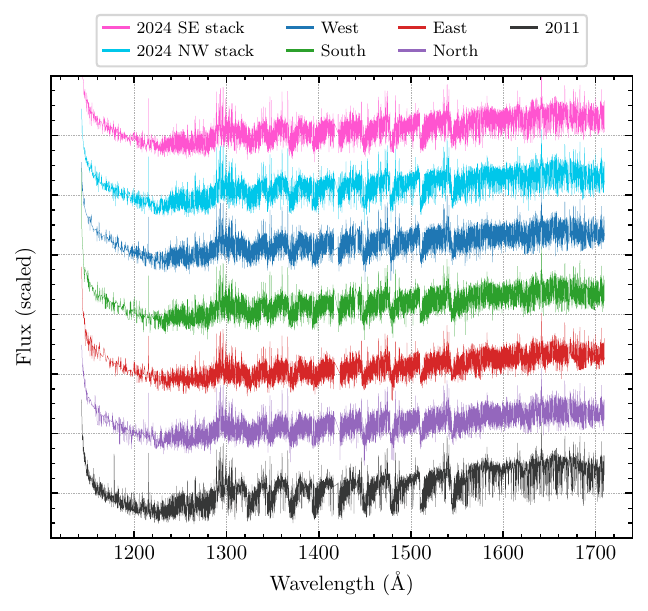}
  \caption{FUV spectra of Betelgeuse from HST STIS E140M measurements. Our four quadrants from the November 2024 observations are shown in purple (N), red (E), green (S), and blue (W), matching the colors in Figure \ref{fig:tiling}. We also show the combined NW and SE `hemispheres' in cyan and magenta. The combined 2011 spectrum \citep{Ayres2010,Carpenter.2018.ASTRALBetel} is indicated by the black curve.}
  \label{fig:fullspec}
\end{figure*}

Both the LSP phase and pulsation phase are important to interpret relative velocity and flux signatures in the UV spectra. As discussed above, the timing of our observations was targeted to match the receding quadrature of $\bbud$, near Betelgeuse's RV minimum on the LSP timescale, approximately halfway between the maximum and minimum brightness phases of the LSP. 

We compare the times of the HST observations to Betelgeuse's light variations in Figure~\ref{fig:hst_times}. The two panels show 2000\,d-long segments, i.e., covering about one LSP cycle with one intermediate segment between the two omitted from the plot. 
The 2011 observations analyzed by \citet{Carpenter.2018.ASTRALBetel}, labeled Ayres+2010 for the observing proposal \citep{Ayres2010}, occurred near the LSP's luminosity maximum on the LSP timescale, but while the pulsation was before and around minimum light. 
In contrast, our new observations \citep{Joyce2024} were taken during the descending-luminosity phase of the LSP (broad behavior in the lower panel of Figure~\ref{fig:hst_times}, see also Figure~\ref{fig:hst_times_rv}) and on the ascending-luminosity phase of the pulsation (local behavior in the lower panel of Figure~\ref{fig:hst_times}), just before maximum light. 

Figure~\ref{fig:hst_times_rv} shows the timing of these same observations relative to the variations in Betelgeuse's RV.
The 2011 observations coincided with the STELLA RV observations, although the RV data are rather sparse at that time. We can subtract the average velocity and a sine fit approximating the 2170-day LSP variation from the STELLA RV data to estimate the contribution from stellar pulsations to the photospheric velocity. We find near-zero to low positive RV excess values that increase over time during the 2011 observations, indicating that the star was near maximum expansion, and then begun to shrink (increasing the redshift of the stellar photosphere, as can be seen in the upper panel of Figure~\ref{fig:hst_times_rv}). Unfortunately, there are no simultaneous RV data publicly available for the new measurements, but we can infer the pulsation phase from the lightcurve. The star's pulsation phase during our observations corresponds to the ascending branch of the light curve (seen in the lower panel of Figure~\ref{fig:hst_times}). The pulsation of the star is currently dominated by the O1 mode \citep{MacLeod2023}, for which \citet{Goldberg2024} found a phase lag between the photometry and the RV close to $-\pi/2$. This, and comparison with earlier simultaneous observations, suggest that the star itself was likely shrinking and approaching maximum compression at the time of our observations \citep{Joyce2024}. 
Therefore, we expect Betelgeuse to be somewhat cooler at the photosphere during the 2011 observations and hotter during our measurements, due to its underlying pulsational behavior.\footnote{Interestingly, the extremely high-velocity compression following the great dimming in April 2020 did not lead to significantly hotter $\Teff$, possibly because of the mass 
ejection (see, e.g. \citealt{Wasatonic2022} who show a long baseline of $V$, TiO strength, and $\Teff$, as well as discussions by \citealt{Levesque2020}).
However, in general the pulsation phase does track the $\Teff$, seen also in the \citet[][]{Wasatonic2022} data.}
Likewise, Betelgeuse's V-band magnitude is slightly fainter during the \citet{Joyce2024} observations as compared to \citet{Ayres2010} ($V\approx0.6$\,mag as compared to $V\approx0.5$\,mag).

\section{Analysis of Observations}
\label{sec:analysis}

We now discuss the similarities and differences between spectra taken in the four quadrants in the context of the search for the putative companion. 
For each of the four quadrants, the two exposures per visit were stacked using the \texttt{stistools}\footnote{\url{https://stistools.readthedocs.io/en/latest/}} and the HASP \texttt{swrapper} script \citep{HASP}. Flux values with SNR$\leq$1 were then masked out in all figures showing spectra. For consistency, we apply the same stacking and masking procedure to the \citet{Ayres2010} observations analyzed by \citet{Carpenter.2018.ASTRALBetel}, and the same masking procedure to the ULLYSES data which we redshift to the companion's expected location as in Figure~\ref{fig:expectations}.

\subsection{Comparison of the 4 quadrants}

\begin{figure}
    \centering
    \includegraphics[width=\columnwidth]{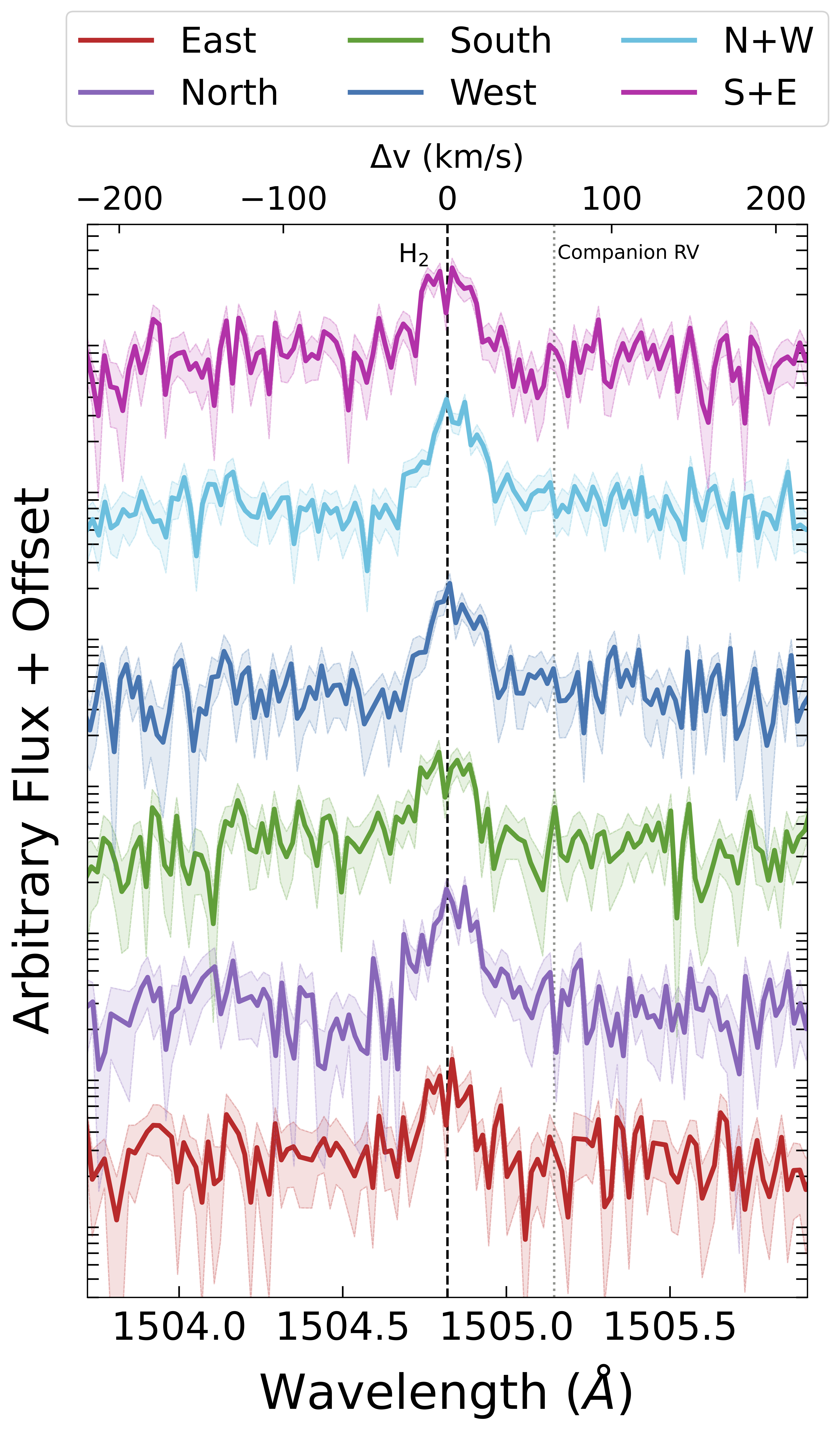}
    \caption{Individual features in all four quadrants, as well as the two combined NW and SE hemispheres, of our FUV spectra of Betelgeuse, centered on the H$_2$ 1504.82\AA\ line (black dashed line). Colors of the quadrants and hemispheres are the same as in Figure \ref{fig:fullspec}. Shaded regions indicate the measurement uncertainty. The location of H$_2$ offset to the companion's expected RV at the time of observations is indicated by a vertical grey dotted line.
    }
    \label{fig:FourQuadrants}
\end{figure}

Figure~\ref{fig:fullspec} shows our complete observations in all four quadrants with the same colors as in Figure~\ref{fig:tiling}, labeled by their approximate relative position on the sky. We also show the stacked north-west (NW) and south-east (SE) `hemispheres', also with the same colors as Figure~\ref{fig:tiling}. 
Each line is offset for legibility. The 2011 observations presented by \citet{Carpenter.2018.ASTRALBetel} are shown in black. Broadly, we see no systematic difference above the noise level between any of the four quadrants. 
Notable emission features have approximately the same width and height in all quadrants, and are comparable to the same features in the 2011 spectrum. The 2011 spectrum is slightly higher in flux at longer wavelengths, consistent with expectations that the stellar surface is slightly cooler due to the Betelgeuse's underlying pulsational phase, and is less noisy than our observations due to the stacking of 6 overlapping exposures compared to our 2 exposures per quadrant (4 per hemisphere). 

Individual spectral features in the different quadrants do exhibit some differences from each other. Figure~\ref{fig:FourQuadrants} shows the region surrounding the H$_2$ 1504.82\AA\  line and measurement uncertainties for the four quadrants as well as the two stacked hemispheres.
Differences in the line shape coming from Betelgeuse itself are to be expected, and have been observed in other spatially resolved spectroscopic observations with HST \citep[e.g.][]{Uitenbroek1998,Dupree2005,Dupree-2020,Granzer2022}. Large-scale convection is expected to be ubiquitous in RSG envelopes \citep[e.g.][]{Stothers1972,Goldberg2022a,Chiavassa2011b,Chiavassa2024,Ma2025}, will induce large-scale asymmetries in measurements of the stellar photocenter \citep{Chiavassa2020} and spatially resolved RV measurements \citep{Ma2024}. If the convective activity seeds chromospheric activity, this should indeed impact Betelgeuse's emission line profiles at the quadrant-by-quadrant and hemisphere-by-hemisphere level. 

Figure~\ref{fig:FourQuadrants} also shows spectroscopic differences at the location of the H$_2$ line at the companion's expected RV, indicated by a vertical grey dotted line. While hints of a feature may appear in some quadrants, there is no clear evidence of emission lines from the companion at its expected RV offset within the measurement uncertainties in this feature. We now turn to discussing other stronger diagnostic lines.

\subsection{Emission from diagnostic lines}\label{subsec:lines_other}

\begin{figure*}
\centering
\includegraphics[width=0.9\textwidth]{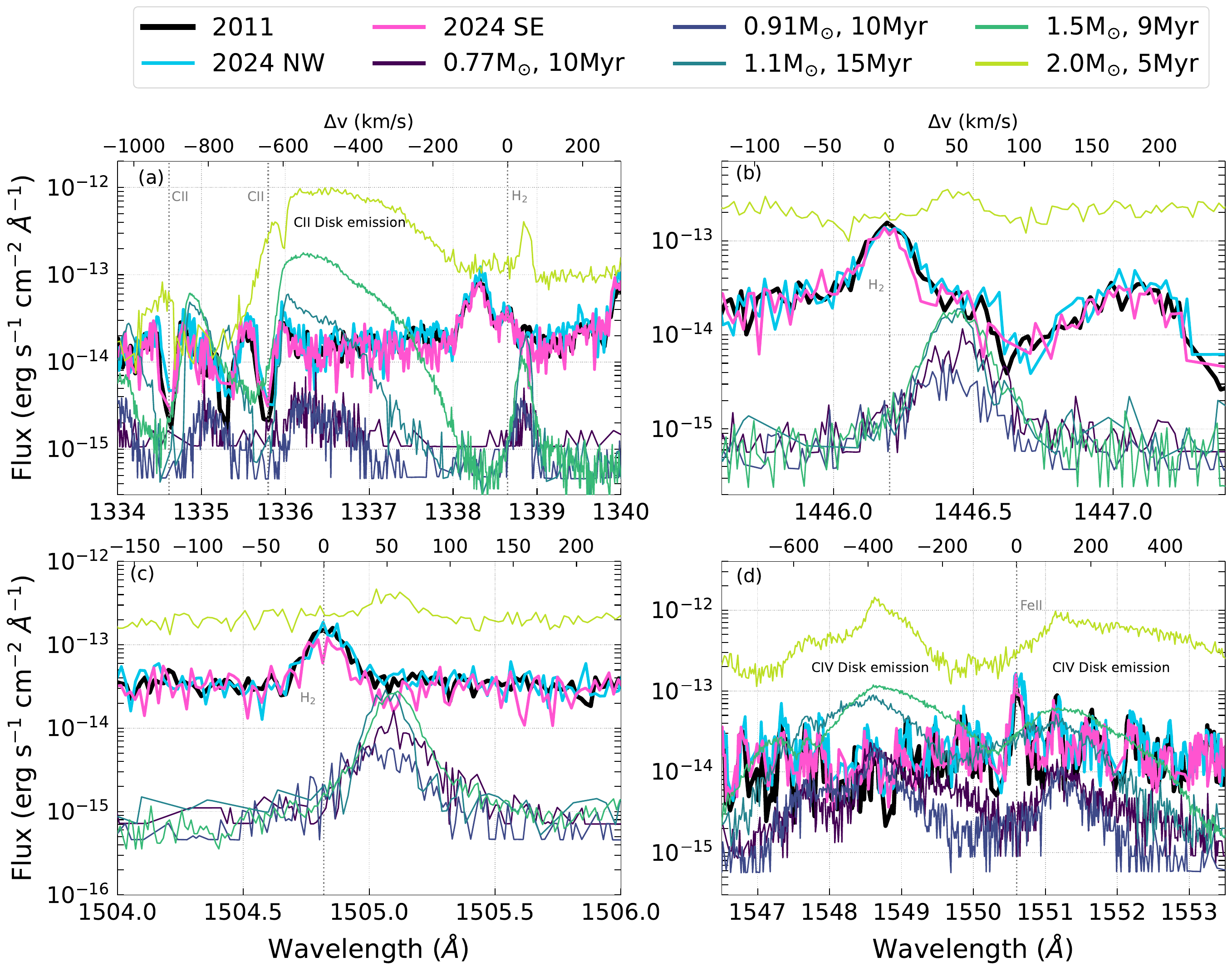}
\caption{HST spectroscopy from our 2024 Betelgeuse observations (NW stack in cyan, SE in magenta), the 2011 ASTRAL observations of Betelgeuse Doppler shifted to Betelgeuse's peculiar+orbital velocity (black), and YSO spectra at a range of ages and masses from ULLYSES, Doppler shifted to the RV of the companion at receding quadrature (purple-blue-green). Panels correspond to the features of interest highlighted in Figure~\ref{fig:expectations} and discussed in \S\ref{sec:expectations}. Wide emission features from disk emission in the ULLYSES data are labeled with text and are not expected to appear in $\bbud$'s emission spectrum (see discussion in \S\ref{sec:fluxexpectations}). 
No substantial difference is observed between the NW and SE observations aside from a slight (few km/s) overall RV shift between them, nor does the 2024 data differ significantly from the 2011 data (during which the relative RVs of Betelgeuse and $\bbud$ would have been minimally offset from each other). } \label{fig:lines_other}
\end{figure*}

While Ly-$\alpha$ normally would be the most promising region to search for emission from a companion, the FUV spectrum shortward of 1235\AA\ in Betelgeuse appears to show no flux above the noise level, potentially absorbed by Ly-$alpha$ damping wings in the outflowing wind. We note also that the feature near Ly-$\alpha$ in Betelgeuse's spectrum is primarily geocoronal (discussed in Appendix~\ref{subsec:lymanalpha}). Given the varied nature of supergiant wind models (see \citealt{vanLoon2025} for a recent review), it is difficult to assess the neutral absorption column along the line of sight to the putative companion's orbital position. Thus we cannot determine whether the observed lack of emission in the Ly-$\alpha$ region is due to absence of a YSO companion, or whether tangential absorption from Betelgeuse's wind is significant enough to mask the presence of a companion with a relatively narrow Ly-alpha emission feature.

Therefore, we turn to alternative wavelength regions containing promising diagnostic lines for inferring the presence of a companion (see Fig.~\ref{fig:expectations} and surrounding discussion in \S~\ref{sec:fluxexpectations}). 
Examples are shown in Fig.~\ref{fig:lines_other}.
We plot the NW (cyan) and SE (magenta) stacked hemispheres of our observations\footnote{We show halves instead of quadrants for ease in viewing, but the results are the same when considering all four quadrants}, the \citet{Carpenter.2018.ASTRALBetel} data (black), and several observations of YSOs from ULLYSES \citet{RomanDuval.J.2022.ULLYSES,ULLYSES2025} at ages and masses indicated in the legend (dark navy-to-green). These YSOs have been distance (168 pc \citealt{Joyce2020})\footnote{At the farther astrometric distance of 222pc \citep{Harper2017}, this would lead to a factor of $1.75$ reduction in flux.} and extinction ($R_V=4.2$ and $A_V=0.65$, \citealt{Montarges2021}) corrected to the parameters of the Betelgeuse system, as detailed in \S\ref{sec:fluxexpectations}.

We note that caution is needed when comparing the observed extinction applied here ($A_V=0.65$) to the discussion of the absorbing Hydrogen column ($N_H$) by \citet{BetelChandra2025}. 
Because the X-ray absorption column close to the stellar surface is unconstrained from direct X-ray observations, \citet{BetelChandra2025} adopt estimates from stellar wind models of $N_H\sim6\times10^{21}-3\times10^{23}\,\mathrm{g}\,\mathrm{cm^{-2}}$ near the companion's location. Even higher estimates of $N_H$ can be inferred from ALMA observations, e.g., integrating the $n_H$ profile determined in recent work by \citet{Dent2024}.
Applying the \citet{Predehl1995} linear relationship between extinction and Hydrogen column for interstellar dust, $A_V=N_H/\left(1.8\times10^{21}\mathrm{\,g\,cm^{-2}}\right)$, 
to even the low ($6\times10^{21}\mathrm{\,g\,cm^{-2}}$) or moderate ($6\times10^{22}\mathrm{\,g\,cm^{-2}}$) values of $N_H$ from \citet{BetelChandra2025} yields implausibly high extinction values of $A_V\approx3$ or $\approx33$, respectively.
In fact, the ISM relation does not apply to the partially-ionized circumstellar regime, where the dust-to-gas ratio is significantly smaller than it is in the farther-out ISM. 
A large hydrogen column in the star's circumstellar wind environment could contribute to photoelectric absorption of X-rays, whereas a smaller amount of dust further out could modestly affect the FUV extinction.


We compare the shapes of emission features between the NW and SE hemispheres, and to the previous 2011 \citet{Carpenter.2018.ASTRALBetel} Betelgeuse observations. These three spectra represent three different system configurations: i) with emission only from Betelgeuse (likely the NW half), ii) with emission from the companion at a maximal RV shift of $\approx$45.5 km/s (likely the SE half), and iii) with emission from the companion at the \textit{same} RV shift (transit) as Betelgeuse \citep{Carpenter.2018.ASTRALBetel}. We find \emph{no} significant differences between the NW and SE halves. Further, with the \citet{Carpenter.2018.ASTRALBetel} data shifted to be at the approximate RV of our Betelgeuse observations, we find no differences between our data and that of \citet{Carpenter.2018.ASTRALBetel}. All significant spectral features are the same to within noise limits. 
We thus report a clear non-detection of blueshifted emission features in excess of the chromospheric flux of Betelgeuse coming from a companion. 

Given this, we look to the ULLYSES YSO data to illustrate what upper limits we can place on the mass and properties of the companion. 
We rule out the presence of a M$\geq$2M$_{\odot}$ companion (light green line), which has flux greater than that of Betelgeuse in all cases. For the 0.77M$_{\odot}$ and 0.91M$_{\odot}$ YSOs (navy lines), even their chromospheric features with the highest flux relative to that of Betelgeuse (panel d) appear too low in flux to be discerned from noise in our spectra.  

For the 1.1M$_{\odot}$ and 1.5M$_{\odot}$ YSOs, the presence of an RV-shifted feature in just one of our observed quadrants/halves would be difficult to disentangle from the noise (see, e.g. the CIV feature in panel d, as well as the CII doublet at 1334.5 and 1335.7 in panel a). 
In panels a), b), and c), for the narrow emission lines and not the disk emission features, we see that the flux of the 1.5$\Msun$ YSO (e.g. the line near 1338.75\AA\ in panel a), or both that and the 1.1$\Msun$ YSO (the CII lines near 1334.5-1336\AA, H$_2$ near 1505\AA), meet or even slightly exceed that of all the Betelgeuse spectra. 

If $\bbud$ is a $\approx$1.1-1.5M$_\odot$ YSO, and has a spectral profile similar to the example YSOs from ULLYSES, then we would rule out those masses, though the diversity in emission from YSOs renders these ULLYSES spectra imperfect templates for the spectrum of the companion. 
Moreover, $\bbud$ will be embedded within the dusty circumstellar material surrounding Betelgeuse, which may obscure or otherwise reprocess emission from the companion. 
Rather, these individual objects are illustrative at best, and their masses estimates. 

While our observational campaign was designed to be maximally sensitive to a possible detection, given the diversity inherent in YSO spectra in the FUV, a non-detection is unfortunately only weakly constraining. 
Take, for example, the strong difference in overall flux in emission features between the 0.91M$_{\odot}$ and 1.1M$_{\odot}$ YSO spectra. Differences of up to an order of magnitude are visible (e.g. panel a around 1336.5\AA), despite only 0.2M$_{\odot}$ between the mass estimates. 
Therefore, we cannot conclusively rule out the presence of a $\sim$1M$_{\odot}$ companion, though a smaller (M$\lesssim$1M$_{\odot}$) is favored by this analysis.

\begin{figure*}
\centering
\includegraphics[width=\textwidth]{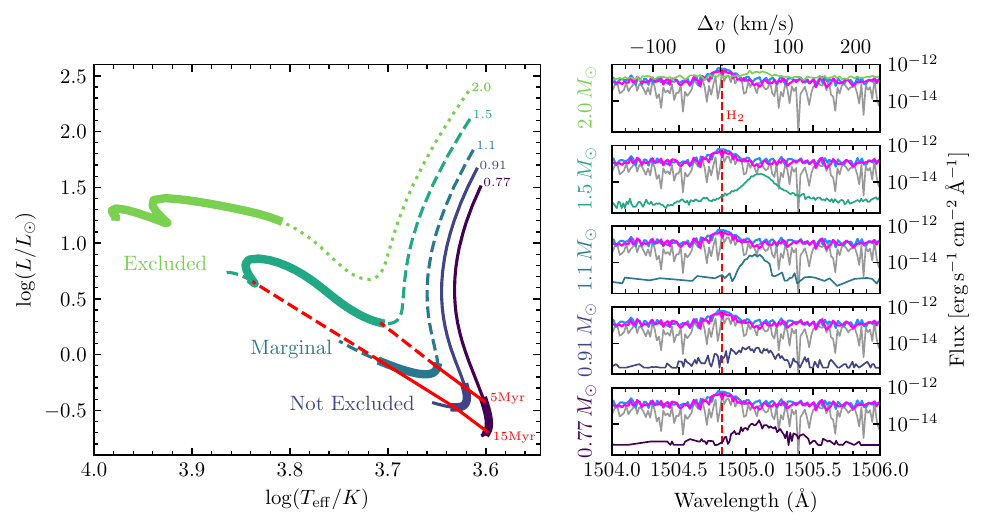}
  \caption{\textbf{Left:} Hertzsprung-Russell Diagram displaying evolutionary tracks from MIST, labeled by mass in $\Msun$, to illustrate excluded and allowable YSO scenarios given the lack of significant spectroscopic differences near the expected redshift of $\bbud$. The plotted tracks span the Hayashi track through truncation at 20 Myr, twice Betelgeuse's approximate age. Thick curves illustrate approximate plausible ages of $\bbud$ ($\approx5-15$Myr), and the red lines are interpolated tracks at fixed age (isochrones) encompassing the 10 Myr age of Betelgeuse with large uncertainties. \textbf{Right:} Individual ULLYSES YSO spectra from panel (c) of Figure~\ref{fig:lines_other} highilighting the 1504.9\AA\ (rest) $ H_2$ feature, ordered by descending mass and adjusted to the companion's redshift, compared to the NW (blue) and SE (magenta) hemispheres of Betelgeuse. The difference between hemispheres, shown in grey, is dominated by a $\approx4$km/s difference which persists across spectral features. The $H_2$ feature (1504.9\AA\ rest frame, adjusted to the approximate redshift of Betelgeuse) is indicated by a vertical red dashed line, and the topmost x-axis indicates the velocity shift relative to Betelgeuse, with $\bbud$ expected to be at $\approx$+45 km/s.}
  \label{fig:hrd}
\end{figure*}

\subsection{Implications for the Companion}
Based on our comparison of observations of the Betelgeuse system during this campaign to characteristic spectral features from YSOs with mass determinations in the ULLYSES database, we
confidently exclude a mass of $2M_\odot$ for a YSO companion. We do not definitively exclude a mass of $1.5 M_\odot$, but find the likelihood to be marginal. Masses below $1.1 M_\odot$ are not excluded. Figure \ref{fig:hrd} illustrates these scenarios with stellar evolutionary tracks (left panel) from MESA Isocrones and Stellar Tracks \citep[MIST;][left panel]{Choi2016,Dotter2016,Paxton2011,Paxton2013,Paxton2015,Paxton2018} at the masses given for the ULLYSES objects shown in Figure~\ref{fig:lines_other}. The tracks shown span the Hayashi track through truncation at twice Betelgeuse's approximate age, 20 Myr (prior to the zero-age main sequence in most cases). Lines of constant age (isochrones, red) at 5 Myr and 15 Myr are shown, with thick curves between (purple-blue-green) liberally indicating approximate plausible ages of $\bbud$. The region bounded by the red isochrones and purple-blue-green evolutionary tracks thus approximately illustrates the marginal (dashed) or fully consistent (solid) remaining locus of the companion (with redder and fainter companions than those shown also fully consistent). 

The mass resolution of the ULLYSES observations we compare to, and more importantly, the extreme diversity of YSO luminosity in UV and high-energy emission (\citealt{Preibisch2005,Preibisch2005Review,Getman2005COUP0} and seen in our Figure~\ref{fig:max}),
make it difficult to place strong constraints on the companion mass. 
Nevertheless, it is interesting that we only encounter true exclusion for a relatively high mass of $\approx2M_\odot$ (dotted green curve), and marginal constraints for the 1.1--1.5$M_\odot$ cases (dashed blue-green curves). 
The fact that a YSO as massive as $1.5M_\odot$ is not fully excluded leaves open 
the need for future resolved and pan-chromatic observing campaigns to place more detailed constraints on the companion's physical nature and interaction with its environment, or, to identify a different cause of the Long Secondary Period persistent in Betelgeuse's photometry, astrometry, and radial velocity from which the putative companion was inferred.

Likewise, flux-calibrated libraries like XSL \citep{Chen2014,Manara2017} extended to the FUV would be valuable for constraining quiet or near-MS companions in the $1.1-1.3\Msun$ range which aren't well-captured in the ULLYSES data set, as broader libraries across a wider parameter space and absolute flux calibration could facilitate a more realistic detection threshold.
Either way, a promising observational path towards tighter constraints on the companion's mass and surface proerties remains long-baseline observations and analysis of historical observations of the resolved stellar surface and circumstellar envelope, specifically informed by timing relative to the LSP phase. 
Likewise, future and proposed FUV missions 
such as the Large UV/Optical/IR Surveyor (LUVOIR, \citealt{TheLUVOIRTeam2019}) and the Cosmological Advanced Survey Telescope for Optical and ultraviolet Research (CASTOR, \citealt{Cote2012}) may provide additional, deeper constraints, in particular on the UV continuum emission of Betelgeuse as well as higher-contrast features from the companion at favorable orbital configurations.

While this work was in the refereeing process, \cite{Howell2025} reported the probable claimed detection, via speckle imaging, of a source 6 magnitudes ($\approx250\times$) fainter than Betelgeuse in the B band, with location and timing predicted by \citet{Goldberg2024} and \cite{MacLeod2025}.\footnote{Sadly, the measurement errors in HST (seen in Figure~\ref{fig:FourQuadrants} and the right panel of Figure~\ref{fig:hrd}) are far too large to observe a $<1\%$ differences in UV continuum flux across quadrants.} 
The authors identify the companion as an $\approx1.5\Msun$ young stellar object approaching but not yet on its Main Sequence. 

This scenario is prima facie at the margin of the range of YSO masses which we can exclude on the basis of a spectroscopic non-detection. However, chromospheric activity typically decreases as stars age closer to the Main Sequence \citep{Preibisch2005Review,Linsky.J.2017.ChromosphereReview}.
Additionally, for a YSO companion interacting with Betelgeuse, spin-down processes may reduce the spin-controlled magnetic activity to below that of a typical pre-MS star. This may impact the emission limits on possible companion types. 
For example, since we expect a YSO companion to be stripped of any residual natal disk as discussed in \S\ref{sec:fluxexpectations}, that may reduce any disk-locking that would otherwise spin up the companion prior to the disk-free T-Tauri stage. Similarly, tidal coupling between Betelgeuse and $\bbud$ as discussed by \citet{MacLeod2025} would reduce the spin of the companion, as Betelgeuse's spin period is slower than the companion's orbital period. These scenarios both may suppress the companion's spin and thus expected UV emission.

Thus, the specific scenario discussed by \citet{Howell2025} is indeed cleanly within the constraints placed of our observations with HST. 
We therefore eagerly await upcoming efforts towards direct detection with speckle imaging, adaptive optics, and other high-resolution high-contrast observational techniques, in order to confirm the anticipated non-detection at occultation and detection at approaching quadrature.

\section{Conclusions}\label{sec:conclusions}
In this work, we detailed our \textit{HST} STIS E140M observations of Betelgeuse during mid-to-late November 2024, near its recently-predicted \citep{Goldberg2024,MacLeod2025} companion's receding limb passage. This resulted in a non-detection:
we do not find any statistically significant line emission in the FUV which could be identified as originating from the companion at the expected RV offset from Betelgeuse ($>+40$km/s redshift, corresponding to a $\approx9$AU orbit with a $\approx2100$d period), within flux limits of $\approx10^{-14}$ erg\,s$^{-1}$\,cm$^{-2}$\,\AA$^{-1}$ which increase with increasing wavelength (across the full band, the average flux is $2.6\times10^{-14}$erg\,s$^{-1}$\,cm$^{-2}$\,\AA$^{-1}$, and the error is $8.4\times10^{-15}$erg\,s$^{-1}$\,cm$^{-2}$\,\AA$^{-1}$). 
There is likewise no statistically significant contribution to continuum emission from a companion (i.e. the continuum differences between quadrants are consistent with HST's error), and continuum-like emission from Betelgeuse is just barely at the $3-\sigma$ level compared to the error. 
Comparing to a set of YSO spectra taken with STIS in the ULLYSES database, we determine that higher-mass scenarios for the companion are ruled out, placing a spectroscopic mass constraint of $M\ltapprox1.5M_\odot$ and favoring masses $M\ltapprox1M_\odot$, with acknowledgment of the great diversity in YSO spectra even within a given mass or spectral-type bin. 
This remains consistent with the lower-mass end of estimates fit from the STELLA RV only \citep{Goldberg2024}, but favors masses in the $0.5-1.1\Msun$ range as predicted by the long-baseline RV analysis \citep{MacLeod2025}. 

Although we cannot uniquely identify spectroscopic features from the companion itself, we see slight differences in the spectra for each quadrant of our HST observations which loosely correspond on the sky to the Northern, Southern, Eastern, and Western quadrants of Betelgeuse's chromosphere and circumstellar envelope. 
Though these differences are only slightly above the background, they likely indicate spatial variation across the chromosphere and circumstellar envelope. This is expected given the scale of the vigorously convective nature of RSG envelopes (\citealt[e.g.][]{Stothers1972,Goldberg2022a,Chiavassa2009,Chiavassa2011b,Chiavassa2024,Ma2025}). These asymmetries are also consistent with resolved interferometric observations of Betelgeuse \citep[e.g.][]{Harper2001, OGorman2015,Montarges-2014, Montarges2021,Kervella2009, Kervella2018, Haubois-2009, Haubois2019, Haubois2023,Kravchenko2021},
as well as other work on spatially resolved UV observations of Betelgeuse and its chromosphere \citep[e.g.][]{Gilliland1996,Uitenbroek1998,Lobel-2001,Harper2006,Dupree2005,Dupree-2020,Matthews2022}.
Further modeling and future spatially-resolved observations may connect spectroscopic differences to spatial differences in chromospheric emission from Betelgeuse. 

Additional X-ray observations with \textit{Chandra}, presented in \citet{BetelChandra2025}, exclude a compact object (white dwarf or neutron star), and constrain the X-ray emission coming from a possible YSO companion with larger uncertainties arising from uncertainty in the X-ray absorption along the line-of-sight as well as the multiple-orders-of-magnitude differences in X-ray emission from YSOs even of the same spectral class. 
Therefore, future observational campaigns remain necessary for direct detection of $\bbud$, in order to confirm its presence and place constraints on its spectroscopic nature, interaction with the stellar wind, and any possible connection to dust formation such as during Great Dimming-like episodes. 
Finally, we re-emphasize the importance of considering the stellar variability cycles, and in particular Long Secondary Periods, when constructing spectroscopic campaigns for Betelgeuse and for all other binary and variable stars. 

\vspace{-0.25cm}
\section*{Data Availability}
All \textit{HST} data used in this work can be obtained from the MAST archive at \dataset[doi:10.17909/wpad-4s76]{http://dx.doi.org/10.17909/wpad-4s76}.

\acknowledgements
\section*{Acknowledgements}
We thank Morgan MacLeod for invaluable discussions. We likewise acknowledge helpful conversations with Andrea Antoni, Matteo Cantiello, Thavisha Dharmawardena, Zarina Dhillon, Kareem El-Badry, Brian Metzger, Michael Palumbo, Adiv Paradise, Mathieu Renzo, and Lieke van Son. We are forever grateful to Bill Paxton for all he has done for the astrophysical community.

J.A.G. is supported by a Flatiron Research Fellowship. The Flatiron Institute is supported by the Simons Foundation.
A.O. and B.O. gratefully acknowledge support from the McWilliams Postdoctoral Fellowship in the McWilliams Center for Cosmology and Astrophysics at Carnegie Mellon University.
M.J. gratefully acknowledges funding of MATISSE: \textit{Measuring Ages Through Isochrones, Seismology, and Stellar Evolution}, awarded through the European Commission's Widening Fellowship. This project has received funding from the European Union's Horizon 2020 research and innovation programme. This research was supported by the `SeismoLab' KKP-137523 \'Elvonal grant and by the NKFIH excellence grant TKP2021-NKTA-64 of the Hungarian Research, Development and Innovation Office (NKFIH). M.R.D. acknowledges support from the NSERC through grant RGPIN-2019-06186, the Ontario Ministry of Colleges and Universities through grant ER22-17-164, the Canada Research Chairs Program, and the Dunlap Institute at the University of Toronto.  

We acknowledge with thanks the variable star observations from the AAVSO International Database contributed by observers worldwide and used in this research. This work has made use of data from the European Space Agency (ESA) mission
{\it Gaia} (\url{https://www.cosmos.esa.int/gaia}), processed by the {\it Gaia}
Data Processing and Analysis Consortium (DPAC,
\url{https://www.cosmos.esa.int/web/gaia/dpac/consortium}). This research made use of NASA’s Astrophysics Data System Bibliographic Services, the \emph{pathfinder} framework \citep{Pathfinder} as well as of the SIMBAD database and the cross-match service operated at CDS, Strasbourg, France. 

\bibliographystyle{aasjournal}
\singlespace

\bibliography{RSG3D.bib, LSP_ADS.bib, BetelBuddy.bib, bib.bib, Chandra_bib.bib, Joyce.bib}

\appendix

\section{Nondetection of Lyman-$\alpha$ Emission from companion}\label{subsec:lymanalpha}

\begin{figure*}
\centering
\includegraphics[width=0.9\textwidth]{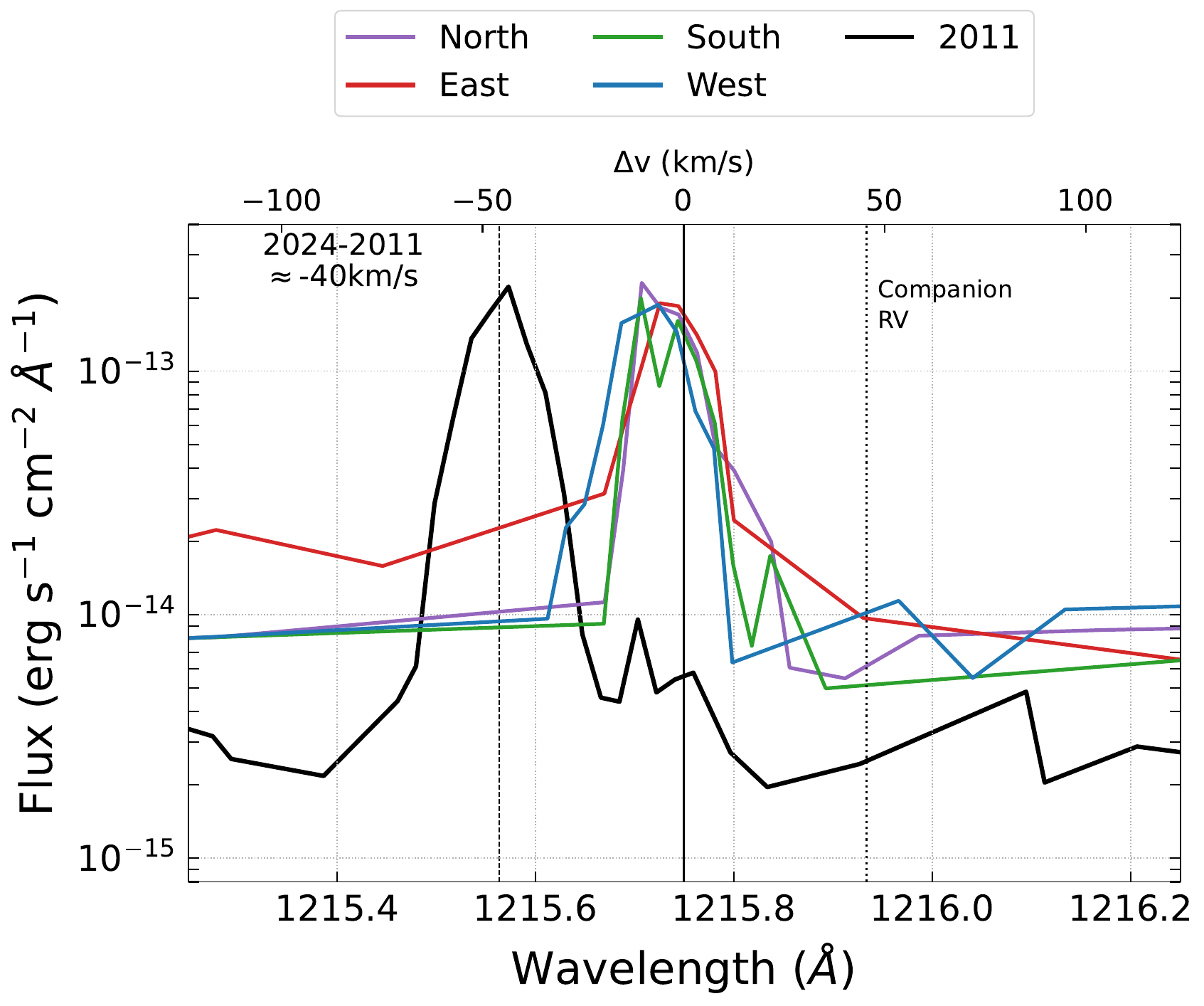}
  \caption{FUV spectra of Betelgeuse centered on the (geocoronal) Ly-$\alpha$ line. Colored lines show the four quadrants of our 2024 observations (colored as in Figure~\ref{fig:tiling}), compared to the 2011 ASTRAL data (black). Vertical lines indicate: Ly-$\alpha$ red-shifted by 19.5 km/s to Betelgeuse's peculiar+orbital RV (solid), and by 65 km/s to $\bbud$'s expected RV (dotted). The center of the 2011 ASTRAL data is indicated with a vertical dashed line. The RV shift between the 2024 and 2011 mean is indicated in text in the upper left corner.
  }
  \label{fig:lines_lyman}
\end{figure*}

The Ly-$\alpha$ (1215.67\AA) emission line is another often-shown feature in the FUV \citep{Linsky.J.2017.ChromosphereReview,Carpenter.2018.ASTRALBetel,RomanDuval.J.2022.ULLYSES,ULLYSES2025}.
However, this feature is dominated by contamination from geocoronal emission from Earth's atmosphere.
We attempted, unsuccessfully, to separate possible source and geocoronal Ly-$\alpha$ emission profiles by extracting cross-dispersion slices of the 2D spectra, including regions centered on the order traces where Ly-$\alpha$ appears along with adjacent background regions on either side. 
Ly-$\alpha$ appears in two adjacent orders in our E140M spectra. For one of the orders, the center trace spectrum shows evidence of a roughly $+$2.5 km s$^{-1}$ shift relative to the background Ly-$\alpha$ spectrum, but no such signal was detected in the other order.

Nonetheless, the companion's expected radial velocity shift is larger than the approximate width ($\approx$35 km/s) of the geocoronal Ly-$\alpha$ line, so if the companion were strongly emitting in Ly-$\alpha$ it could in-principle be visible here. 
Therefore, for completeness, Figure~\ref{fig:lines_lyman} shows spectra near the Ly-$\alpha$ feature from each of the four observed quadrants, along with the 2011 observations by \citet{Carpenter.2018.ASTRALBetel} shifted to the RV of Betelgeuse at the time of our 2024 observations. There is no discernable Ly-$\alpha$ emission at the expected redshift of the companion.

Due to the geocoronal origin of the strong Ly-$\alpha$ features here, the $\sim40$ km/s offset between the 2011 Ly-$\alpha$ and the median 2024 Ly-$\alpha$ can be attributed to differences in velocity about the Earth's orbit projected toward Betelgeuse between the different epochs. 
The broad nature of the Ly-$\alpha$ features is likewise attributed to the diffuse sky-glow filling the 0.2x0.2 aperture.
Interestingly, Ly-$\alpha$ is the only emission feature throughout the entire spectrum that displays this large of a velocity offset between the 2024 and 2011 observations; 
all other features are only offset by $\approx$2.5 km/s between the 2011 and our 2024 observations, owing to the RV difference in Betelgeuse between these dates. This includes the 1302-1306\AA\ OI lines, which are also contaminated by geocoronal emission.

We also compared with template spectra (not shown here) of YSO's from ULYSSES \citep{RomanDuval.J.2022.ULLYSES,ULLYSES2025}, but those spectra are likewise contaminated by geocoronal emission. 
Without clean expectations for the uncontaminated Ly-$\alpha$ emission profiles of possible companion scenarios, it is unfortunately impossible to place mass constraints from the lack of observed Ly-$\alpha$ from the companion.

\end{document}